\documentclass{mn2e}
\input epsf.sty

\def\hi{\chi^2_{\nu}}
\def\NH{N_{\rm H}}
\def\NHEBV{N_{\rm H, E(B-V)}}
\def\nH{n_{\rm H}}
\def\Rg{R_{\rm g}}

\def\Rin{R_{\rm in}}
\def\Ptot{P_{\rm tot}}

\def\kT{k T_{\rm e}}
\def\taue{\tau_{\rm e}}

\def\Or{\Omega/2\pi}
\def\Gamsoft{\Gamma_{\rm soft}}
\def\LEdd{L_{\rm Edd}}
\def\LX{L_{\rm X}}
\def\Eedge{E_{\rm edge}}
\def\Ftot{F_{\rm tot}}
\def\Fbb{F_{\rm bb}}
\def\Fdbb{F_{\rm discbb}}
\def\Fhard{F_{\rm hard}}
\def\Fseed{F_{\rm seed}}
\def\Fcompt{F_{\rm compt}}
\def\Tic{T_{\rm IC}}
\def\Tin{T_{\rm in}}
\def\Teff{T_{\rm eff}}
\def\Tcol{T_{\rm col}}
\def\uE{u_{\rm E}}
\def\me{m_{\rm e}}

\def\tedge{\tau_{\rm edge}}
\def\kappaabs{\kappa_{\rm abs}}
\def\kappaes{\kappa_{\rm es}}
\def\sigmaKN{\sigma_{\rm KN}}
\def\MEdd{\dot M_{\rm Edd}}
\def\mdot{\dot m}
\def\MSun{{\rm M}_{\odot}}
\def\ginga{{\it Ginga}\ }
\def\xte{{\it RXTE}\ }
\def\OmegaK{\Omega_{\rm K}}
\def\Gammas{\Gamma_{\rm soft}}
\def\tentoto{10^{21}\,{\rm cm^{-2}}}

\def\ergs{{\rm erg\ cm^{-2}\ s^{-1}}}

\title[Soft spectral states of Soft X-ray Transients]
{On the complex disc--corona interactions in the soft spectral states
of Soft X-ray Transients}

\author[P. T. \.{Z}ycki, C. Done and D. A. Smith]
         {Piotr T. \.{Z}ycki$^1$\thanks{E-mail: ptz@camk.edu.pl}, 
          Chris Done$^2$ and  David A. Smith$^{3}$ \\
    $^1$ Nicolaus Copernicus Astronomical Center, Bartycka 18, 00-716 Warsaw,
Poland, \\
    $^2$ Department of Physics, University of Durham,
        South Road, Durham DH1 3LE \\
    $^3$ Department of Astronomy, University of Maryland, College Park, 
   MD 20742-2421, USA}

\date{3 June 2001}

\begin{document}

\maketitle

\begin{abstract}

Accreting black holes show a complex and diverse behaviour in their
soft spectral states. Although these spectra are dominated by a soft,
thermal component which almost certainly arises from an accretion
disc, there is also a hard X--ray tail indicating that some fraction
of the accretion power is instead dissipated in hot, optically thin
coronal material. During such states, best observed in the early
outburst of Soft X--ray Transients, the ratio of power dissipated in
the hot corona to that in the disc can vary from $\sim 0$ 
(pure disc accretion) to $\sim 1$ (equal power in each).
Here we present results of spectral analyses of a number of sources, 
demonstrating the presence of complex features in their energy spectra. 
Our main findings are: (1) the soft components are not properly described
by a thermal emission from accretion discs: they are appreciably
broader than can be described by disc blackbody models even including
relativistic effects,  and (2) the spectral
features near 5--9 keV commonly seen in such spectra can be well described
by reprocessing of hard X--rays by optically thick,  highly ionized, 
relativistically moving plasma.

\end{abstract}


\section{Introduction}

Black Hole Soft X-ray Transient (SXT) sources are a sub-class of 
Low Mass X-ray Binaries occasionally undergoing dramatic outbursts. 
From a dim quiescent state
they brighten (in the opt/UV/X/$\gamma$-ray bands) by several orders 
of magnitude 
in the course of a few days, and then decline, with roughly exponential
dependence of X-ray flux on time (see Tanaka \& Lewin 1995; Tanaka \&
Shibazaki 1996; Chen, Shrader \& Livio 1997 for reviews). The outbursts
are thought to be caused by a sudden increase of mass accretion rate
onto the central object as a result of the disc instability due to
hydrogen ionization. The basic mechanism is thus the same as
in dwarf novae (e.g. Osaki 1996), 
but with important differences due to larger and 
more massive accretion discs and the intense X-ray irradiation in SXT
(Cannizzo 1993; King \& Ritter 1998).

Black Hole SXT are perfect laboratories to study the accretion process.
Black holes are simpler accretors than neutron stars or white dwarfs
thanks to the lack of global magnetic field and a hard surface. 
During an outburst and following decline the mass accretion rate changes 
by several orders of magnitude, giving data on how the accretion processes
change as a function of $\dot m$.
Results from SXT studies can then be applied to other systems with accretion
onto black holes, e.g.\ active galactic nuclei (AGN). If similar outbursts
take place in AGN, they cannot be observed in progress due to much longer
(factor of $10^5$--$10^7$) time scales (Burderi, King, \& Szuszkiewicz
1998; Siemiginowska, Czerny, \& Kostyunin 1996).

Previous investigations of both black hole and neutron star SXT led to 
identification of a number of distinct spectral/temporal states
(see e.g. Tanaka \& Lewin 1995; van der Klis 1995 for reviews): (1) quiescence
state, when the sources are very dim ($\LX \sim 10^{33}\ {\rm erg\ s^{-1}}$),
with (poorly constrained) power law spectra, (2) low/hard state,
(LS) characterized by a hard, roughly power law spectrum (photon index 
$\Gamma \sim 1.7$) and a strong X-ray variability (r.m.s.\ 20--30 per cent),
with power spectral density (PSD) peaking at $\sim 0.1$ Hz; 
(3) intermediate state (IS), when
the power law component is somewhat softer ($\Gamma\sim 2$) and a weak soft, 
thermal component appears in the spectrum; (4) high/soft (HS) state with
X-ray spectra dominated by a thermal component of temperature $\sim 1$ keV
and PSD of rather low amplitude (a few per cent) and a power law shape;
(5) very high state (VHS) when both the soft, thermal ($k T \sim 1$ keV) and 
the hard power law components ($\Gamma =  2-3$) are
present in the spectrum, and similarly the PSD is a mixture of PSDs
characteristic to LS and HS. Additionally, quasi-periodic oscillations
of various frequency (1--10 Hz) and strength appear in PSD in this state.
For the purpose of this work the IS, HS and VHS are referred to as soft
states, since it is in these states that a strong soft, thermal component
is present in the spectra.
Both the directly observed fluxes and temporal sequence of occurrence of these
states during SXT's decline leave little doubt that the accretion rate is 
lowest in quiescence and increasing through LS, IS, HS to VHS. 

A general scenario for the evolution of accretion flow in SXT as a
function of accretion rate was presented by Esin, McClintock \&
Narayan (1997).  They suggested that the most characteristic feature
of the evolution during the decline -- the high to low state
transition -- involves a major change of the flow geometry. An
optically thick, thermally emitting accretion disc dominant in soft
states is replaced by a hot, optically thin(ish) flow able to produce
the strong, hard X-ray emission seen in the low state by thermal
Comptonization (e.g.\ Gierlinski et al.\ 1997). The transition
presumably involves evaporation of the optically thick disc (e.g.\
Meyer \& Meyer-Hofmeister 1994; R\'{o}\.{z}a\'{n}ska \& Czerny 2000),
whose inner radius increases from the last stable orbit at $6\,\Rg$ in
the high state to $\sim 10^4\,\Rg$ in the low state ($\Rg\equiv G
M/c^2$ is the gravitational radius). 

One way to track the optically thick accretion disc is through
reflection of the hard X--ray spectrum by the accretion disc.  The
reflected component consists of the Compton reflected continuum with
spectral features due to iron K shell absorption and fluorescence
imprinted on it (Lightman \& White 1988; George \& Fabian 1991; Matt,
Perola \& Piro 1991).  When the reprocessing matter forms a disc
rotating in a deep potential well, the spectral features are broadened
and smeared by Doppler effects and gravitational redshift (Fabian et
al.\ 1989; Ross, Fabian \& Brandt 1996). The observed spectral
features between 6--9 keV can be fit by such models and the derived
amount of reflection and relativistic smearing is consistent with the
disc inner edge moving inwards as the LS makes a transition to the
IS/HS/VHS (\.{Z}ycki, Done \& Smith 1998; Gilfanov, Churazov \&
Revnivtsev 2000). However, the amount of reflection and smearing seen in
the LS generally precludes the disc being as far away as $10^4\, \Rg$
as in the simplest versions of the truncated disc models (Esin et al.,
1997), and instead requires that the optically thick material extends
down to $\sim 30\, \Rg$ (\.{Z}ycki, Done \& Smith 1997; 1998; 1999a;
Miller et al.\ 2001). While a (somewhat modified) truncated disc model
can then fit the LS data, this is not a unique interpretation: the disc
could extend down to the last stable orbit if there is complex
ionization structure in the disc and/or relativistic outflow of the
hard X--ray emitting plasma (see Done 2001 for review and the
discussion in Di Salvo et al.\ 2001).

At higher luminosities, in the IS/HS/VHS the spectrum is dominated by
the soft component from the accretion disc (e.g.\ Ebisawa et al.\ 1994)
as expected in the Esin et al.\ (1997) model, but this co--exists 
(especially in the VHS) with a hard X--ray tail. This tail 
is probably produced by non--thermal Comptonization (Coppi 1999; 
Gierli\'{n}ski et al.\ 1999) as it can extend without a break to a
few hundred keV (Grove et al.\ 1998).  The ratio of luminosities in the
hard and soft components can assume a broad range of values, from
$\sim 0$ where the spectrum is pure disc emission to $\sim 1$ with
equal luminosities in the disc and X--ray hot phase (Ebisawa et al.\
1994; Homan et al.\ 2001; Rutledge et al.\ 1999; Sobczak et al.\ 
2000). As well as the continuum, iron spectral features at 5--10 keV
are again usually observed in the spectra (Ebisawa et al.\ 1994). They are
interpreted as indicating reprocessing of hard X--ray radiation by an
optically thick accretion disc, but detailed reflection models are not
often fit to the data. A common spectral model is of a disc blackbody
(Mitsuda et al.\ 1984)
and power law continuum with a broad iron line and smeared edge to
phenomenologically model the reflected spectrum from the disc.
While this model gives a rough idea of the ratio of soft to hard luminosity, 
it has some 
obvious drawbacks. Firstly, the disc spectrum presumably
provides the seed photons for the Compton upscattering, so the hard
spectrum cannot extend as an unbroken power law into the disc blackbody
spectrum. A power law continuum overestimates the contribution of the
hard continuum at soft energies, and so this suppresses and distorts
the derived soft component.  Also, a true reflected spectrum contains
a continuum as well as spectral features. Neglecting the continuum distorts the
derived power law spectral index.  The few papers which have modelled
the reflected spectrum show unambiguously that the reprocessing medium
in SXT in the IS/HS and VHS is highly ionized (\.{Z}ycki et al.\ 1998;
Miller et al.\ 2001; Wilson \& Done 2001), unlike the LS where it is
mainly neutral. Smearing of the reflected spectral features is
observed in all spectral states (\.{Z}ycki et al.\ 1997, 1998, 1999a;
Miller et al.\ 2001; Wilson \& Done 2001). 

If the physics of the accretion flow really is controlled
predominantly by $\dot{m}$, then the SXT spectra should be the same as
those in the persistent X--ray sources at similar accretion rates. The
most thoroughly studied soft state so far -- in Cyg X--1
(Gierli\'{n}ski et al.\ 1999) -- shows all the above listed features:
non-thermal Comptonization as the origin of the hard X--rays, and the
ionized and smeared reprocessed component.  Additionally, it
conclusively shows that the continuum is not well modelled by a single
disc blackbody spectrum which is Comptonized by a non--thermal
electron distribution. Additional low temperature Comptonization of
the disc spectrum is {\it required\/} by the data (Gierli\'{n}ski et
al.\ 1999; Coppi 1999).

In this paper we re-analyse some of the archival \ginga and \xte data
of SXTs in the soft states. We show that in most cases the spectra
{\it cannot} be described by a physically motivated model of a simple
(disc)blackbody emission, with a Comptonized power law tail which is
reflected by the accretion disc. Similarly to the HS Cyg X--1 spectrum
described above, the soft component {\it requires\/} a broader spectral
form than that of a disc blackbody (even including relativistic
corrections). This could be due to additional thermal Comptonization
in the disc, or to a separate blackbody component from X--ray heated
hot spots on the disc. The poor fit of the disc blackbody model is
contrary to most results derived from phenomenological models of the
hard component and spectral features (power law plus broad line and/or
smeared edge) which generally give adequate fits to the data, and on
the few occasions where they did not (e.g.\ Ebisawa et al.\ 1994), then
the evident problems in the modeling made any inference about the
soft component shape unconvincing.  The phenomenological models for
the spectral features have more free parameters than the physically
motivated reflected spectral models, and so can compensate for
difficulties in the continuum model.  We demonstrate that when the
spectral features at 6--9 keV are modelled as X--ray reprocessing, the
reprocessor is strongly ionized, as expected for a disc at temperature
$\ge 0.5$ keV.

\section{Data and Models}

We analyse \ginga Large Area Counter (LAC) data of GS~1124-68 
(Nova Muscae 1991) and
GS~2000+25.  {\it Ginga\/} data are still some of the best to study
spectra with strong soft components, thanks to its broad band energy
coverage from $\sim 1-20$ keV.  The data were re-extracted from LEDAS
public archive at Leicester University. The background was subtracted
according to the 'universal' procedure based on blank sky observations
(Hayashida et al.\ 1989). Whenever the source counts contaminated the
Surplus above the Upper Discriminator (SUD) monitor, we first
recovered the original SUD values using procedure described in
\.{Z}ycki et al.\ (1999a), and then applied the 'universal' background
subtraction. We assume 0.5 per cent systematic errors in the data. 
This is motivated by the uncertainty in LAC response from fits to the
Crab spectrum (Turner et al.\ 1989), as well as the residual uncertainty
in the determination of the LAC background (Hayashida et al.\ 1989).

We note that for data from instruments like {\it Ginga}/LAC
or {\it RXTE}/PCA the systematic error is often the dominant error up
to about 10 keV. Its value then has a direct influence on the
$\hi$ values obtained. While this makes the absolute value of
$\hi$ and the confidence contours on fit parameters lose their
usual statistical meaning, the $\hi$
values do preserve the ranking of models.
Since some of our best fits have $\hi$ substantially 
less than 1, it suggests that the assumed systematic error may be
somewhat overestimated.

The \xte data (PCA top layer only, detectors 0, 1 and 2 only) 
were extracted from public archive at HEASARC/GSFC, and
reduced with {\sc FTOOLS} ver.\ 4.2 software. The background was estimated
employing the latest background models appropriate for bright sources
(the 'sky-VLE' model for Epoch 3 of \xte observations).  Systematic
errors of 1 per cent were assumed in each data channel.

The \ginga data we model here were previously analysed by Ebisawa
(1991) and Ebisawa et al.\ (1994). They used the phenomenological
model of a disc blackbody (Mitsuda et al.\ 1984) plus a power law with
smeared edge and broad gaussian line near 6-8 keV. Here we model the
spectral features using the {\sc rel-repr} model (\.{Z}ycki et al.\
1999a). The model uses the angle-dependent Green's functions for the 
problem of Compton reflection as computed by Magdziarz \& Zdziarski (1995;
as implemented in XSPEC models {\sc pexrav}/{\sc pexriv}), to find
the reflected continuum for a given primary continuum. The Fe
K$_{\alpha}$ line is computed as in \.{Z}ycki \& Czerny (1994), with
photo-ionization computations as in  Done et al.\ (1992), and is then
added to the reflected continuum, so that
the properties of the reflected continuum and the line
are computed consistently for a given ionization state, inclination
and elemental abundances.  Relativistic and kinematic corrections are
then applied by convolving the spectrum with the the appropriate
Green's function for Schwarzschild metric, using the prescription of
Fabian et al.\ (1989). The final model thus have three main
parameters: (1) the ionization parameter, $\xi\equiv 4\pi F/\nH$,
where $F$ is the irradiating flux in 5 eV -- 19 keV band and $\nH$ is
the hydrogen number density, (2) the overall amplitude (normalized
to the underlying continuum), which is
equivalent to the solid angle of the reprocessor as visible from the
X--ray source normalized to $2\pi$, $\Omega/(2\pi)$ and (3) the inner
disc radius, $\Rin$ determining the relativistic smearing of the
spectral features (the irradiation emissivity is assumed $\propto
r^{-3}$, \.{Z}ycki et al.\ 1999a). All radii are given in 
units of the  gravitational radius ($\Rg = G M/c^2$).

We note that {\sc pexriv} and {\sc rel-repr} are based on computations
of irradiated discs with constant density. In more advanced computations
with hydrostatic equilibrium, the thermal instability of X--ray irradiated
plasma plays a role, and leads to a more complex vertical structure,
with important observational consequences
(R\'{o}\.{z}a\'{n}ska \& Czerny 1996; Nayakshin, Kazanas \& Kallman 2000).
However, the thermal instability is probably unimportant for the soft state
spectra in SXT's as 
the thermal equilibrium curve (the ionization parameter --
temperature, $\Xi$--$T$ diagram) degenerates since its lower stable branch
is at high disc temperature $T_0 \sim 0.5-1$ keV. The soft spectrum
dominates the spectrum, so the Compton temperature is 
not much higher than $T_0$ and the instability is suppressed.

We use two analytical models approximating the inverse-Compton
spectra: {\sc thComp}, based on solution of the Kompaneets equation
(Zdziarski, Johnson \& Magdziarz 1996) and 
{\sc comptt} (Titarchuk 1994). We  check
whether our conclusions may be affected by the approximate character
of the Comptonization 
models, using the {\sc compPS} Comptonization model (Poutanen \&
Svensson 1996), which finds a numerical solution of the Comptonization problem
for optically thin ($\taue \le 1$) plasma explicitly considering 
successive scattering orders. This code can be used with a variety
of electron energy distributions, including a purely thermal distribution,
a purely non-thermal one or a hybrid electron distribution, i.e.\ 
a Maxwellian distribution with a power law tail above certain Lorenz factor,
$\gamma_1$ (Coppi 1999). Such a hybrid electron distribution can
provide a good description of X--ray/$\gamma$-ray
spectra of Cyg X--1 in the soft state (Gierli\'{n}ski et al.\ 1999)
as Compton scattering of the disc  blackbody by 
the thermal part of the electron distribution provides an additional,
broad soft component while the 
X--ray power law is produced by Compton scattering of the disc photons
by the non--thermal electron tail.

The soft component will be modelled by the {\sc diskbb} model (Mitsuda
et al.\ 1984). We will also use a general relativistic disc spectrum model
{\sc grad}, which integrates black body spectra over disc radius, taking
into account: the colour temperature correction (constant with radius)
at $\Tcol/\Teff= 1.7$ (Shimura \& Takahara 1995),
fully relativistic radial dependence of disc temperature in Schwarzschild 
metric (Kato, Fukue \& Mineshige 1998) and kinematic/relativistic effects 
on photons propagation,
using the method of Fabian et al.\ (1989). A similar model for the Kerr
metric, with the black hole spin $a=0.998$  will be also used. Here we
employ the radial effective temperature prescription from Kato et al.\ 
(1998), the colour temperature correction constant with radius, 
while the photon propagation is done by convolving the local spectra
with the transfer function of Laor (1991), as implemented in {\sc XSPEC}.
We note that the colour temperature correction to a blackbody spectrum 
is appropriate when the escaping photons undergo saturated Comptonization.

We use {\sc XSPEC} ver.\ 10 for all spectral fits
(Arnaud 1996), with the above non--standard models implemented as local
models. Photoelectric absorption is modelled using 
{\sc wabs} (Morrison \& McCammon 1983), and parameter errors are given
as $\Delta\chi^2 = 2.7$.

\section{Intermediate State data}

First, we analyse a number of datasets of various sources obtained when 
the sources were
about to make a transition from high state to low state. We refer to them
as Intermediate State data.

\subsection{GS~2000+25}
\label{sec:gs2000}

\begin{table*}
 \caption{Results of model fitting of the Dec 8th spectrum (IS) 
of GS~2000+25}
 \label{tab:GS2000_1}
   \begin{tabular}{llcccc}
parameter &   units   &      A          &       B         & 
                                        C            &        D     \\
\hline
$\NH$     & $\tentoto$& $0^{+0.6}$      &     6.0(f)            &
                                    $0^{+0.7}$     &    6.0(f)        \\
\hline
$T_0$     & keV       & $0.32 \pm 0.01$ &    $0.18\pm 0.1$         &
                                  $0.32\pm 0.01$  &   $0.18\pm 0.01$    \\
$\kT$     & keV       &    --           &     10(f)             &
                                      --         &   10(f)         \\
$\Gamsoft$&           &    --           & $7.1\pm 0.3$            &
                                      --        &   $7.3 \pm 0.3$        \\
\hline
$\Eedge$   & keV       & $8.1^{+0.7}_{-0.6}$ &  $7.8\pm 0.3$       &
                                     --         &     --               \\
$\tedge$   &           & $0.52 \pm 0.28$   &    $32 \pm 9$         &
                                     --          &    --             \\
\hline
$\Gamma$  &           & $2.39 \pm 0.07$    &   $1.93\pm 0.15$       &
                       $2.54^{+0.07}_{-0.10}$  &  $2.30^{+0.11}_{-0.13}$  \\
\hline
$\Or$     &           & --              &   --    &
                            $0.44 \pm 0.32 $  &   $0.36^{+0.24}_{-0.13}$    \\
$\xi$     & erg cm s$^{-1}$& --              & --          & 
                     $50^{+\infty}_{-25}$ & $(1^{+30}_{-0.8})\times 10^3$\\
$\Rin$    &  $\Rg$    & --              &  --  &
                 $ > 20$       &     $>8 $         \\
\hline     
$\hi$     &           & 38/23           &   17.9/23        &
                           30.0/23        &   10.6/23               \\
\hline
   \end{tabular}

Model A: absorption(diskbb + smedge(powerlaw))

Model B: absorption(thComp + smedge(powerlaw))

Model C: absorption(diskbb + thComp + rel-repr)

Model D: absorption(thComp + thComp + rel-repr)

\end{table*}

\begin{table*}
 \caption{Results of model fitting ($\chi^2$/dof) of the Intermediate State 
spectra of GS~2000+25}
 \label{tab:GS2000_2}
   \begin{tabular}{lcccccccc}
dataset &   model  A     &  model   B         & 
                 model    C          & model   D   &
                        model 0   & model E    & model F   & model G \\
\hline
Sep 8th  & 28.9/24       &  11.0/23 & 90.0/24$^{N}$ &  14.2/24$^{f}$   &
                   199/26   &   15.6/25$^f$  &  18.7/23$^f$ & 66.7/24 \\
Oct 18th & 43.9/24       &  14.5/23 & 159/24        &  17.2/23         &
                   314/26   & 21.0/24    &  12.5/22         & 80.7/24 \\
Nov 5th  & 28.5/24       &  14.5/23 & 113/24        &  14.4/23   &
                   404/26    & 19.2/24   &  12.1/22         & 60.6/24  \\
Dec 8th  & 38.0/23$^{N}$ &  17.8/22 & 30.0/23$^{N}$   &  10.6/23$^{f}$   & 
                  46.3/25$^{N}$ & 15.1/25$^f$ & 9.0/22$^f$ &  32/23$^{N}$ \\
\hline
   \end{tabular}

Model A: absorption(diskbb + smedge(powerlaw))

Model B: absorption(thComp + smedge(powerlaw))

Model C: absorption(diskbb + thComp + rel-repr)

Model D: absorption(thComp + thComp + rel-repr)

Model 0: absorption(thComp + thComp), i.e.\ model D without the reprocessed 
component

Model E: absorption(compPS + rel-repr), hybrid Comptonization

Model F: absorption(diskbb + bbody + thComp + rel-repr)

Model G: absorption(grad  + thComp + rel-repr)

$^{N}$ -- best fit $\NH$ significantly lower than the interstellar value

$^{f}$ -- $\NH$ fixed at the interstellar value
\end{table*}

\begin{figure} 
% .qdp 
 \epsfxsize = 0.45\textwidth 
 \epsfbox[60 360 470 720]{dec8_eeuf.ps}
%.sm 
\epsfxsize = 0.45\textwidth 
 \epsfbox[50 250 500 630]{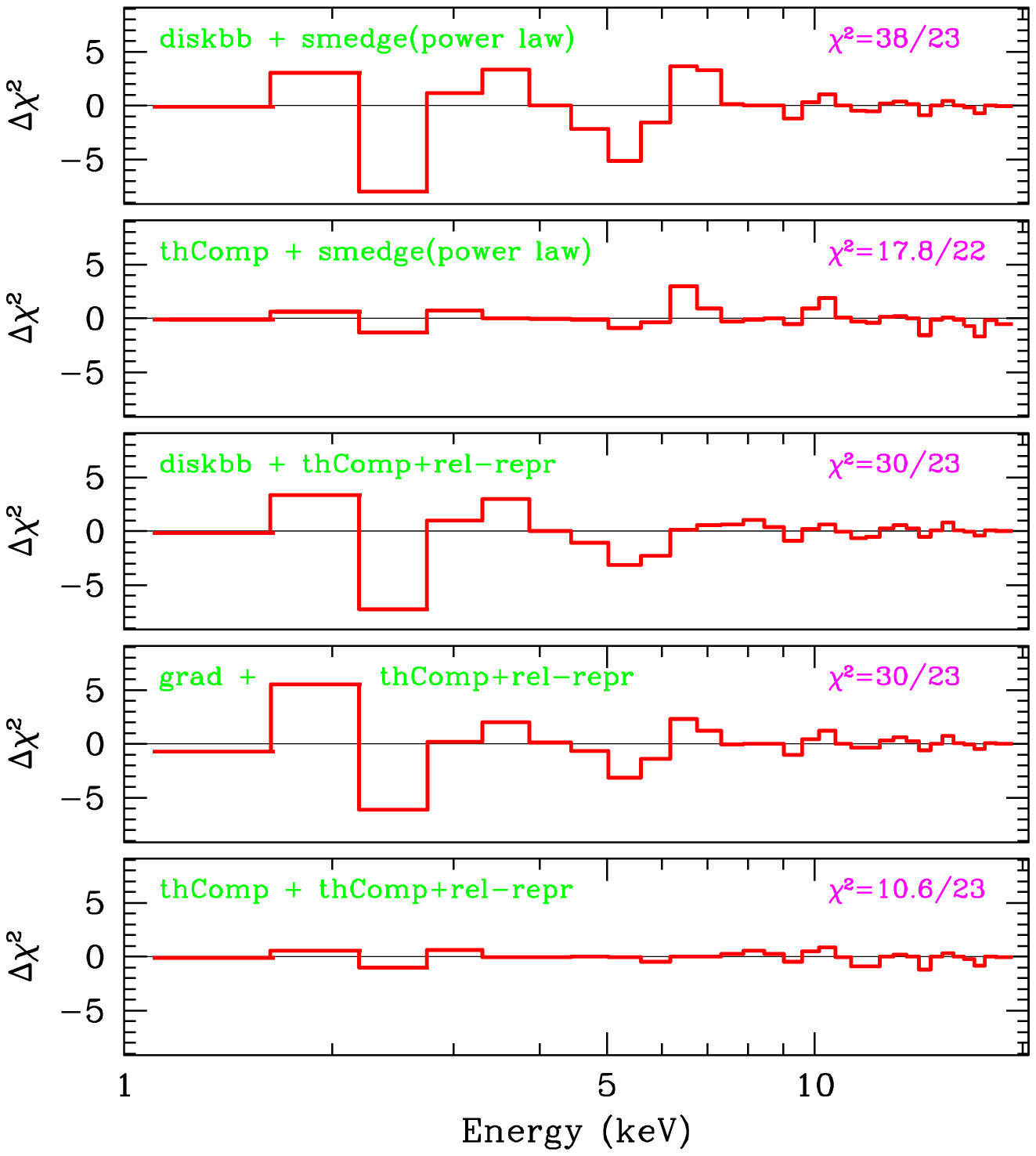} 
\caption{{\it Upper panel\/}:  
The unfolded spectrum of GS~2000+25 on Dec 8 1988, shortly before its 
transition to the
low state, showing a Comptonized soft component, Comptonized hard
power law tail and its
X-ray reflection from the disc (model D in Table~\ref{tab:GS2000_1}). 
{\it Lower panel\/}: contributions to 
$\chi^2$ from different models (Table~\ref{tab:GS2000_1}). 
The best fit model contains 
Comptonized soft component and the physical description of X--ray 
reprocessing: {\sc thComp + rel-repr}. 
\label{fig:gs2000sp}} 
\end{figure}

We first analyse the \ginga 1--20 keV spectra of GS~2000+25, beginning with 
the Dec 8, 1988
dataset (day 224 after the peak on April 28, the last observation in the HS; 
see Ebisawa 1991, Tanaka \& Lewin 1995, fig. 3.11). Repeating the analysis
of Ebisawa (1991), 
we use the {\sc diskbb} model and a power law tail with 
the smeared edge applied only to the latter component,
{\sc wabs*(diskbb + smedge*(powerlaw))} (model A in Table~\ref{tab:GS2000_1}). 
This phenomenological
description gives a rather poor fit, $\hi = 38/23$, 
with best fit $\NH = 0^{+0.7} \times 10^{21}\,{\rm cm^{-2}}$, i.e.\ 
lower than the likely interstellar  value of 
$\approx 6 \times 10^{21}\,{\rm cm^{-2}}$. Adding a Gaussian line to 
the model does not improve the fit irrespectively of whether the line
is narrow or broad, and its energy fixed at 6.4 keV or free.
%dec8_diskbb_smedge.xcm
Replacing the disc blackbody 
model by the Comptonized blackbody ({\sc thComp}) results in significant 
improvement, $\hi = 17.9/23$ (with $\NH$ fixed at the interstellar
value; model B in Table~\ref{tab:GS2000_1}).
%dec8_thcomp_smedge.xcm
The fit is even better when the hard power law with smeared edge is 
replaced by Comptonization model {\sc thComp} (so that the full model
contains two {\sc thComp} components) with its corresponding
reprocessed component, {\sc rel-repr}: the fit has $\hi = 10.6/23$
(model D).
%dec8_thcomp_thcompfe.xcm
As a test, we return to the {\sc diskbb} model for the soft component, but
we keep the {\sc thComp+rel-repr} model for the hard component
(model C in Table~\ref{tab:GS2000_1}).
We obtain a significantly worse fit, with $\hi = 30.0/23$, even though 
$\NH$ was free in this fit ($\chi^2$ increases to 99.7 when $\NH$ is fixed 
at the interstellar value). Thus the best fit is obtained with the 
proper physical description of the Fe spectral features {\it and} a
Comptonized blackbody model for the soft component (model D in
Figure~\ref{fig:gs2000sp}).

A very similar sequence of results is obtained for the earlier
observations of GS~2000+25 (Nov 5th, Oct 18th, Sep 8th, see
Table~\ref{tab:GS2000_2}), when the source luminosity was higher by up
to a factor of $\sim 5$ than on Dec 8th.  In all cases better fits are
obtained with the soft component modelled as Comptonized blackbody than
a disc blackbody, irrespective of the model used for the hard
component (a power law with smeared edge or {\sc thComp} with {\sc
rel-repr}, although the latter is a more physical description of the source
and gives a better fit).  Since the power law slope of the soft
component is rather high (steep spectrum), the Comptonization
temperature, $\kT$, is not well constrained by {\sc thComp}. Thus the
soft component can be well fit by either low temperature, high optical
depth plasma or by high temperature, low optical depth
plasma. However, for the latter case when $\taue\le 1$, the
approximations on which {\sc thComp} is based become invalid. Accurate
Comptonized spectra are not a power law as the individual scattering
orders can be seen, as can the initial seed photons. Using the {\sc
compPS} model (which gives accurate Comptonized spectra including the
seed photons for $\taue \le 1$) for the soft component gives an upper
limit on the temperature of $\approx 25$ keV, 
%sep_compps_thcompfe.log 
corresponding to $\taue \approx 0.55$ for
spherical geometry for the Sep 8th spectrum.  Thus these data favour a
low temperature, optically thick(ish), unsaturated 
($y \equiv 4\kT/(\me c^2)\,\taue \sim 0.1$) 
Comptonization of the soft component.

The {\sc compPS} model can also be used with a hybrid electron
distribution (both thermal and non--thermal electrons). This
hybrid Comptonization model provides a good description of all the
datasets (Model E: Table~\ref{tab:GS2000_2}). The thermal part of the
hybrid electron distribution Comptonizes the disc blackbody photons
into a broader soft component, while the non--thermal electrons
Comptonize the disc photons to make the harder spectrum which is
reflected from the disc.  The parameters of the thermal Comptonizing
plasma are similar in all cases: $\kT \approx 5$ keV, $\taue=1-2$ (for
a spherical geometry). They are consistent with the parameters derived
above for a purely thermal origin for the soft excess ({\sc compPS} or
{\sc thComp}), but are rather different from the parameters derived
from fitting the high state spectrum of Cyg X--1, where $\kT \sim 40$
keV was required (Gierli\'{n}ski et al.\ 1999).

\begin{table}
 \caption{Parameters of the reprocessed component from fitting 
 the double Comptonization model to GS~2000+25 data}
 \label{tab:gs2000_3}
   \begin{tabular}{lccc}
parameter &          Nov 5         &        Oct  18         &        Sep 8   \\
\hline
$\Or$     & $0.71^{+0.73}_{-0.25}$ & $0.52^{+0.68}_{-0.18}$ & 
                                                       $0.43^{+1.4}_{-0.18}$ \\
$\xi$     & $(18^{+30}_{-12})\times 10^3$  &
                                 $(10^{+25}_{-8.4})\times 10^3$ & 
                                             $(13^{+50}_{-12}) \times 10^3$  \\
$\Rin$    & $15^{+13}_{-4.5}$      & $8.2^{+3.1}_{-1.9}$    &  $6^{+3.2}$    \\
\hline
   \end{tabular}
\end{table}

Another possibility to describe the complex soft component in these
spectra is to assume a sum of a disc blackbody and an additional single
blackbody. The resulting fits are good (model F:
Table~\ref{tab:GS2000_2}), with the
disc blackbody temperature at $k T_{\rm dbb} \sim 0.3-0.4$ keV, 
with a higher temperature for the additional blackbody at
$k T_{\rm bb} \sim 0.5-0.8$ keV.

The broadness of the soft component {\it cannot\/} be simply attributed
to relativistic effects: the {\sc grad} model, which includes these effects,
 does not provide a good description of our data. The $\chi^2$ values
shown in Table~\ref{tab:GS2000_2} are usually even worse for the {\sc grad}
model (model G) than for the simple disc blackbody (model C). In these fits the
inner disc radius has been fixed at the marginally stable orbit at $6 \Rg$. 
The fits can be improved if $\Rin$ is a free parameters, in which case 
$\Rin$ usually increases to $\ge 20\,\Rg$. The relativistic effects
are not important then, and the model is equivalent to {\sc diskbb}.

The data clearly require a presence of the reprocessed component. To
illustrate its significance we removed this component from the double
Comptonization model (model D), and fit parameters of the two continua, 
including
$\kT$ for the first component and $\NH$. This resulted in bad fits,
as given as model 0 in Table~\ref{tab:GS2000_2}. The 
derived parameters for the reprocessed
component are similar to those previously
inferred for soft states of black hole systems (\.{Z}ycki et al.\ 1998;
Gierli\'{n}ski et al.\ 1999): the main features are that the reprocessor
is highly ionized and it requires further smearing/broadening compared to
predictions of the simple model. This is demonstrated in 
Table~\ref{tab:gs2000_3}, where we show parameters of the reprocessed 
component, as obtained from fitting model D to the Nov 5, Oct 18 and
Sep 8 datasets. We model the smearing as due to relativistic and kinematic 
effects, but spectral features can be additionally
significantly broadened by Compton down- and upscattering, as the photons 
diffuse through the ionized disc layers (e.g.\ Ross, Fabian \& Young 1999). 
Therefore, $\Rin$ as determined from our fits is likely to represent the
lower limit on the inner radius of the reflecting disc.

The bolometric luminosity of GS~2000+25 is rather low, if the distance
to the source $d=2\pm 1$ kpc  (Callahan et al.\ 1996)  is adopted. 
Even for the earliest dataset analysed, Sep 8th, the bolometric luminosity, 
$L=(1.1-2.0)\times 10^{37}\,d_2^2\,{\rm erg}\ {\rm s^{-1}}$ 
corresponds to less than 2 per cent
of the Eddington luminosity. The luminosity on Dec 8th is 
$\sim 0.003\,\LEdd$, while one week later, during the first observation in the
 LS the luminosity is $\approx 2 \times 10^{-4}\, \LEdd$. 
It is rather unclear whether
this source was showing the same spectral characteristics as 
other sources (both transient and persistent), despite the low $L/\LEdd$,
or the distance is severely underestimated.

\subsection{GS~1124-68}

{\it Ginga\/} spectral data of GS~1124-68 (Nova Muscae 1991) have been 
analysed by a
number of authors (Ebisawa 1991; Ebisawa et al.\ 1994; Esin et al.\ 1997).
We have already analysed a number of datasets in \.{Z}ycki et al.\ (1998).
In particular, on 18-May-1991 ($\approx 150$ days after the peak of the 
outburst) the source was observed in a soft state for the last time. 
Fitting the {\sc wabs(diskbb + thComp + rel-repr)} model
to the 1--20 keV data we obtain a very bad fit, $\hi=241/24$,
with overall absorption column $\NH=0$.
% diskbb_thcompfe.xcm
Replacing the {\sc diskbb} model by Comptonized blackbody, modelled by
a second {\sc thComp} component, yields dramatic improvement, 
$\hi=14.8/24$, with $\NH$ fixed at the interstellar value $1.6\times\tentoto$
%thcomp_thcompfe.xcm
(note that in \.{Z}ycki et al.\ 1998 we used a power law to
approximate the Comptonized hard component). 
Again the Comptonized soft component is rather steep, so although its 
slope is well constrained ($\Gamma_{\rm soft} =5.6 \pm 0.2$), 
$\kT$ and $\taue$ cannot be separately determined in the approximate
{\sc thComp} models. Using {\sc CompPS}
for the soft component again demonstrates that the data require that 
the Comptonizing medium has low temperature and high optical depth. 

The results are robust with respect to the model for the hard component:
the {\sc wabs(diskbb + compPS + rel-refl} model gives 
$\hi = 320/23$, i.e.\ clearly an unacceptable fit.
% diskbb_compps
Replacing the {\sc diskbb} component with {\sc thComp} we again obtain
a very good fit, $15.5/23$. Thus the presence of the additional Comptonization
on the soft component 
does not seem to be an artifact of the approximate character of the
Comptonization models.

The hybrid Comptonization model gives a good fit to the data,
with $\hi = 17.1/25$. The Maxwellian temperature is $\kT = 5.40 \pm 0.05$ keV,
the cloud optical depth is $2.94\pm 0.10$, similar to those
derived for GS~2000+25. The model comprising a disc blackbody with 
an additional blackbody gives $\hi=18.5/23$. The model with the 
{\sc grad} 
component for the soft emission does not give a good fit, $\hi=215/25$.

\subsection{XTE~J1550-564}

The outburst of XTE~J1550-564 in September 1998 and subsequent
evolution were very well covered by pointed \xte observations.  Basic
spectral analyses of the large collected dataset were performed by
Sobczak et al.\ (2000 and references therein), 
while correlations between timing and
spectral properties were studied by Homan et al.\ (2001). Wilson \&  Done
(2001) studied in the source spectral evolution on the rising
phase of the outburst, as the source made a transition from the low
state to the very high state. However, here we are interested in
Intermediate State. We identify this as occurring during the long
minimum after the peak and before the source flux recovered again (see
ASM lightcurve in fig.~1 in Homan et al.\ 2001), and we analyse a
spectrum taken in the middle of this interval, on 22 Nov 1998 (MJD=51139). 
 
\begin{figure} 
% .qdp 
 \epsfxsize = 0.45\textwidth 
 \epsfbox[20 350 500 720]{j15_eeuf.ps} 
%.sm 
\epsfxsize = 0.45\textwidth 
 \epsfbox[50 430 550 720]{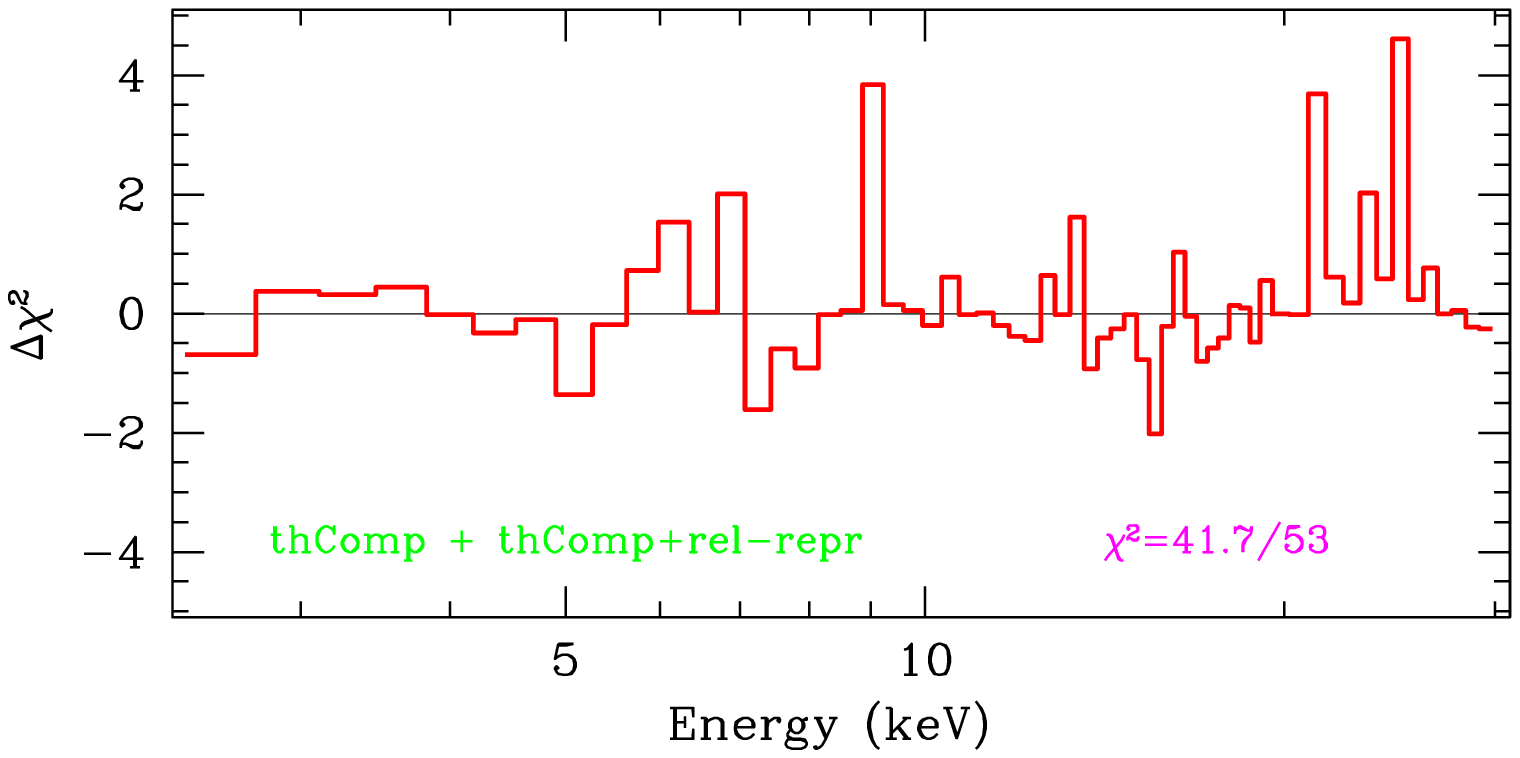} 
\caption{{\it Upper panel\/}:  
Unfolded spectrum of XTE~J1550-564 on Nov 22, 1998, during the long
minimum after the peak of its outburst,
 showing Comptonized soft component, hard power law tail and 
X-ray reprocessed component. {\it Lower panel\/}: contributions to 
$\chi^2$.
\label{fig:j1550_sp}} 
\end{figure} 

We  fit the 2.5--30 keV data with the
{\sc wabs*(diskbb + thComp + rel-repr)} model,
i.e.\ the hard component is modelled as a Comptonized emission
with a corresponding reprocessed component. When $\NH$ is fixed at 
the likely interstellar value of $4\times\tentoto$ (Wilson \& Done 2001),
the fit is poor, $\hi = 79/54$, with the strongest residuals 
% pcu012_top_diskbb_thcompfe.xcm
between 5 and 9 keV.
Allowing for free $\NH$ improves  the fit somewhat; the best fit has 
$\hi = 72/53$ for $\NH = 0$.
% pcu012_top_diskbb_thcompfe_freenh
Substituting now a second {\sc thComp} component in place of the {\sc diskbb}, 
yields a significant improvement, the best fit has $\hi = 41.7/53$
with $\NH$ fixed at $4\times\tentoto$ (Fig.~\ref{fig:j1550_sp}). 
The hard component has a slope of 
$\Gamma_{\rm hard} = 2.12^{+0.05}_{-0.09}$ and its reprocessed
spectrum is detected with 
amplitude $\Or=1.1^{+\infty}_{-0.65}$ from 
highly ionized $\xi = (30^{+65}_{-23})\times 10^4$ material
which is strongly 
smeared, corresponding to $\Rin=6^{+3}\,\Rg$ (assuming the disc inclination 
of $60^{\circ}$). Removing the reprocessed component altogether gives
a very poor fit with $\hi = 185/56 $.
%pcu012_top_thcomp_thcompfe.xcm

%pcu012_top_thcomp_thcompfe.log
The parameters of the {\sc thComp} model for the 
soft component are similar to those derived for GS~2000+25 and GS~1124-68 in
that they give a steep power law tail to the disc blackbody 
($\Gamma_{\rm soft} = 5.4 \pm 0.4$). 
This broadening of the soft component can be equally well matched by 
hybrid Comptonization models ($\hi=46.7/54$) or % pcu012_top_compps_hybrid.xcm
as the sum of disc blackbody with an additional (hotter)
blackbody ($\hi=28.2/52$). The fit with the {\sc grad} model is rather worse,
$\hi=69.5/54$         % pcu012_top_grad_rin6

Adding HEXTE data to the fit does not change our conclusion about
the Comptonization of the soft component. With two {\sc compPS} components
we obtain a good fit, $\hi = 57/87$,  % pcu012_top_hexte_compps_x2
with the parameters of the soft component very similar to the previous
case of PCA data alone.

The distance to the source is rather uncertain, as is the mass estimate 
of the compact object and indeed its very nature.
Estimates of the $L/\LEdd$ ratio are therefore very uncertain.
The 2--30 keV flux in these data is $\approx 5.5\times 10^{-9}\,\ergs$ 
which is $\approx 30$ times lower than
that on 19th Sept (MJD = 51075), during the bright flare. However,
our spectral analysis indicates that significant flux was emitted below
2 keV. The bolometric correction is rather uncertain, depending primarily
on whether the seed photons are assumed to be simple blackbody or disc
blackbody. We derive the bolometric (unabsorbed), 0.1--1000 keV flux
to be
$(1.4 - 2.7) \times 10^{-8}\,\ergs \approx (3-5) \times F(2-30\,{\rm keV})$.
On the other hand, the bolometric
correction for the flare spectrum seems to be $ < 2$, so the ratio
$L_{\rm flare}/L \sim 10$, possibly less.

Adopting the distance $d\approx 2.5$ kpc, based on optical observations 
during the 1998 outburst (S\'{a}nchez-Fern\'{a}ndez et al.\ 1999),
we obtain rather low peak luminosity, 
$L_{\rm flare} \approx 2\times 10^{38}\,\ergs$, hence 
$L\approx 2\times 10^{37}\,\ergs$. The latter corresponds to 
$L/\LEdd \approx 0.03\, d_{2.5}^{2} M_6^{-1}$ (where $M_6 \equiv M/(6\MSun)$),
similar to
the soft--hard transition luminosity for Cyg X--1 
(Gierli\'{n}ski et al.\ 1999; Di Salvo et al.\ 2001).

%------------

\section{The High and Very High State data}

A number of SXTs showed spectral/timing states classified as a VHS.
Nova Muscae 1991 showed such a behaviour  during 
the first $\sim$month around the peak of the outburst (16th Jan 19991)
(Ebisawa et al.\ 1994; Takizawa et al.\ 1997; Belloni et al.\ 1997;
Rutledge et al.\ 1999). GRO~J1655-40 went through VHS during its
outburst in March -- September 1997 (M\'{e}ndez, Belloni \& van der Kils 1997).
XTE~J1550-564 displayed a variety of spectral/timing states including
the VHS (Homan et al.\ 2001; Wilson \& Done 2001). 
A peculiar spectrum was shown by
GS~2023+338 at the peak of its 1989 outburst: the spectrum was very hard,
with high energy cutoff at the low value of $\sim 20$ keV 
(\.{Z}ycki, Done \& Smith 1999b).

\subsection{GS~1124-68}

\subsubsection{VHS -- Jan 11th data}

\begin{figure*}
 \hfil
 \epsfysize = 8 cm
 \epsfbox[20 400 620 680]{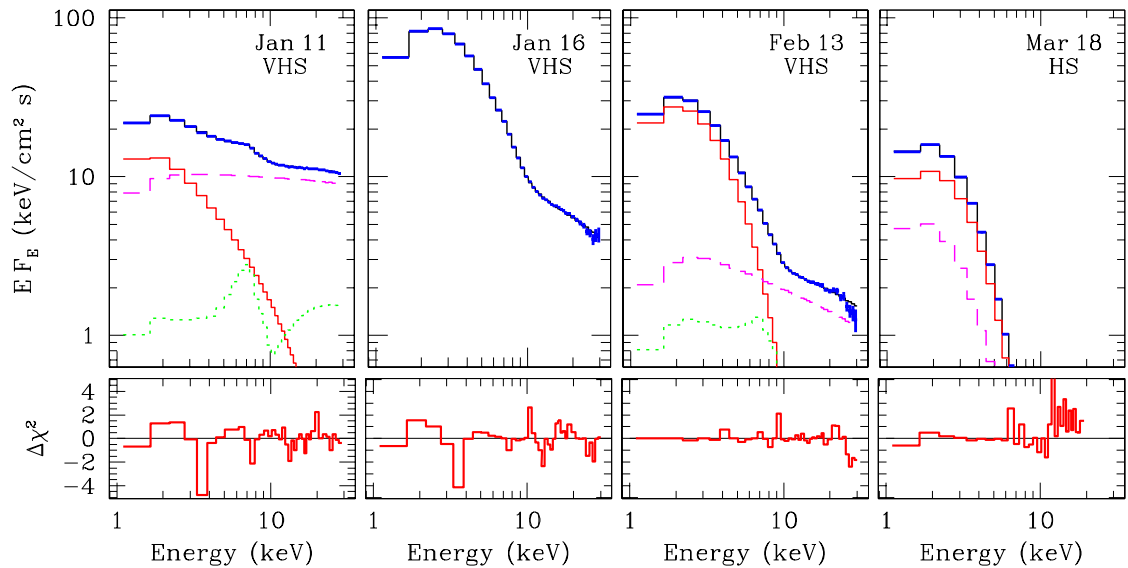}
 \hfil
\caption{Spectra of GS~1124-68 (Nova Muscae 1991) in Very High and High states.
The thick solid histogram represents the best fit total model, the 
histograms drawn in thin line plot the model components: solid histogram -- 
the soft component, dashed histogram -- the hard component, dotted histogram --
the reprocessed component. In all shown spectra the soft component
is significantly better described by a Comptonized (disc) blackbody than
the simple disc blackbody emission.
On Jan 11 the source was still on the rising phase of the outburst. 
The source luminosity peaked on Jan 16 (the hybrid Comptonization model
is plotted for these data). On Feb 13 the bolometric luminosity
is very similar to that on Jan 11, but the two spectra is rather different
(Miyamoto et al.\ 1995). In the last spectrum shown the hard component
is negligible.
\label{fig:VHHS}}
\end{figure*}

During this observation the source was still on the rising phase of 
the outburst, but its bolometric luminosity (with the best model described 
below) was fairly high, $L \approx 0.25 \LEdd \,d_{3.5}^2\,M_6^{-1}$ 
($F=1.4\times 10^{-7}\,\ergs$).
Similarly to the case of the IS spectra, we find here that the soft component 
cannot be described by the disc blackbody model.
More specifically, the best fit
of the {\sc wabs (diskbb + thComp + rel-repr)} model has $\hi = 184/31$.
%diskbb_thcomp.xcm
Replacing the disc blackbody component by the Comptonized blackbody
model {\sc thComp}, gives
us a good fit, $\hi = 27.7/30$. %thcompfe.xcm
This confirms our earlier result (\.{Z}ycki et al.\ 1999b, see caption
to fig.~5), that the soft component in that spectrum is Comptonized,
in addition to the obvious Comptonization of the hard component
(note that our previous result was obtained with a power law model for the 
hard component, while here we use a Comptonized spectrum for it). 
The spectrum is plotted in Fig.~\ref{fig:VHHS}. 
Reflection of the hard component is significantly present in the spectrum, its 
amplitude is $\Or = 0.24^{+0.21}_{-0.03}$,
it is strongly ionized $\xi = (3^{+4}_{-2.7})\times 10^3 $ and 
additionally smeared. The smearing corresponds to inner disc radius of 
$\Rin=16^{+8}_{-4}$. The model with no reprocessed component gives
$\hi=555/33$. The spectral slope
of the soft Comptonized component is $\Gammas = 2.94^{+0.10}_{-0.26}$, 
rather flatter than seen in the IS spectra above. 
With {\sc thComp} or {\sc comptt}, we can only give the lower
limit to the plasma temperature, $\kT> 2$ keV. The soft component can also be 
well described by the sum of disc blackbody ($k T_{\rm dbb} \approx 0.5$ keV) 
and  and additional, hotter ($k T_{\rm bb} \approx 1.2$ keV) blackbody, 
giving the overall
$\hi=13.0/31$. We were not able to find a good fit with the hybrid 
Comptonization model ($\hi \ge 150/32$). Similarly, the {\sc grad} model
used for the soft component gives the overall poor fit with $\hi=239/33$.

Note that the overall spectrum is rather hard ($\Gamma \sim 2$)
and it lacks a dominant soft component, making it rather similar to typical
low state spectra. This led Miyamoto et al.\ (1995) 
to suggest that the transition from spectral hard to soft state occurs
at much higher luminosity on the rising phase of the outburst than
on the declining phase, i.e.\ the source shows a kind of hysteresis.
However, with more detailed spectral fitting it is evident that 
the VHS spectrum
is rather different from typical low/hard state spectra: the soft
component is not a pure disc blackbody emission and the reprocessed features
correspond to highly ionized plasma. On the other hand, during the decline,
spectra at similar luminosity level were totaly dominated by the soft 
component (see discussion below and Fig.~\ref{fig:smdot} and \ref{fig:VHHS},  
the Jan 11.\ vs.\ Feb 13.\ spectrum).
Thus indeed the developement/destruction of the optically thick disc
spectra does not seem to be symmetric in $\dot m$, in a time-dependent
situation.

\subsubsection{VHS -- Jan 16th data}
\label{sec:jan16}

On Jan 16th the source's flux reached maximum. The spectrum is dominated
by a soft thermal component with a weak, steep power law tail
(Ebisawa et al.\ 1994).  

Fitting the 
%jan16_thcompds.xcm
{\sc wabs (diskbb + thComp + rel-repr) } model to 1--30 keV data we
obtain a poor fit, $\hi = 73.8/32$, for $\NH$ fixed at the
interstellar value $1.6\times \tentoto$. Allowing $\NH$ to be free
does not improve the fit. The strongest residuals are below $\sim 5$
keV, again indicating a more complicated shape of the soft component.
Replacing the disc blackbody with a Comptonized component gives an
extremely good fit: {\sc wabs (thComp + thComp + rel-repr) } (with
blackbody seed photons) has $\hi=11.7/32$ (with $\NH$ fixed). 
%thcomp_thcompfe 
However, as opposed to the Intermediate State spectra,
here the curvature of the soft component is clearly seen and,
consequently, the electron temperature is very tightly constrained at
$\kT = 1.12^{+0.04}_{-0.07}$ keV.  The Thomson depth of the
Comptonizing plasma is large, with $\taue \approx 8 $ for slab
geometry. %comptt1_thcompfe.xcm

Again, an acceptable fit can also be obtained by a hybrid electron
distribution: 
{\sc diskbb+ compPS + rel-repr} gives $\hi=28.3/32$ assuming that the seed
photons for the {\sc compPS} are the disc blackbody, % compps_hybrid_freenh.xcm
although this requires $\NH$ larger than the interstellar value, 
$\NH=(3.5 \pm 0.8)\times\tentoto$. The thermal part of the hybrid
electron distribution is then again 
a rather cool, $\kT = 6$ keV, optically thick plasma, $\taue = 2.3$.
Alternatively the soft component can be modelled as a sum
of a simple blackbody and a disc blackbody. This gives $\hi=12.1/30$
% bb_dbb_thcompfe.xcm
and parameters: $k T_{\rm bb} \approx 0.50$ keV and 
$k T_{\rm dbb} \approx 1.1$ keV, thus the blackbody seems to have a lower
temperature than the disc blackbody component, opposite to what was observed
for IS spectra. Energetically, the blackbody component
contributes $\approx 30$ per cent of the total flux of the soft component.
The {\sc grad} model for the soft component does not provide a good
fit, giving the overall $\hi=82.4/33$.

The dominance of the soft component in these spectra give perhaps the
best view of its shape. Hence we have also fit the data a model of
multi-temperature blackbody emission with the radial dependence of the
temperature different from the standard formula, $T(r) \propto
r^{-3/4}$. Our motivation for such a model comes from recent work by
Watarai et al.\ (2000), who draw attention to the fact that in
luminous accretion discs, $\dot M \sim {\dot M}_{\rm Edd}$, the $T(r)$
dependence is flatter than the standard one due to a fraction of
energy being advected radially. We have tried both a simple model,
with $T(r)=\Tin (r/\Rin)^{-\alpha}$ ($\Tin$ and $\alpha$ free
parameters), and a more complex one, where $T(r)$ was actually
computed from the radial structure of a vertically averaged
Shakura--Sunyaev (1973) disc, with the radial advection included.  However,
neither of the two models (together with {\sc thComp + rel-repr} for
the hard component) provides a good description of the data. The
broadening detected in the soft component is much more evident that
can be produced by such models. While it is to be expected that the simple
disc blackbody approximation breaks down when the source luminosity reaches 
$\sim \LEdd$ (see, e.g., Beloborodov 1998), simple
'first order' corrections to the disc blackbody models are not sufficient.

Note that in \.{Z}ycki et al.\ (1999b) we have analysed this dataset and 
concluded
that it is compatible with pure disc blackbody emission with a hard tail.
i.e.\ no additional Comptonization of the soft component was required
to make the fit statistically acceptable.
That conclusion resulted, however, from our using larger systematic 
errors of 1 per cent. More specifically, assuming the 1 per cent 
systematic 
error, the model {\sc diskbb+ thComp + rel-repr} gives  $\hi=33.0/32$.
However, the model {\sc thComp + thComp + rel-repr} gives $\hi=7.5/30$
when fit to the same data. The conclusion that the Comptonized
blackbody model gives  a better fit than the disc blackbody is obviously
independent of the assumed value of systematic errors.

\subsubsection{VHS -- Feb 13th data}
\label{sec:feb13}

This is also a Very High state spectrum, at a luminosity lower
by a factor of $\approx 2$ than the peak luminosity. Here we also find 
evidence for a Comptonized soft component.

The model {\sc wabs(diskbb + thComp + rel-repr)} gives 
$\hi=119.3/25$    % feb13_diskbb_thcompfe.xcm
for fixed $\NH$. 
Letting $\NH$ free gives $\hi=63.8/24$ and $\NH=0^{+0.3}\times\tentoto$,
clearly less than the interstellar value. Replacing {\sc diskbb}
with the relativistic model {\sc grad} improves the fit somewhat, giving
$\hi = 38.8/24$.  % feb13_grad_rin_6_chi39.xcm

The Comptonized blackbody model with a hard tail gives a very good fit,
$\hi = 8.0/23$, using the {\sc thComp} model for the soft component,
or $\hi = 9.8/23$ using the {\sc comptt} model. % feb13_comptt_thcompfe.xcm
Again the data require the additional Comptonizing medium to be
at fairly low temperature $\kT = 0.85^{+0.09}_{-0.05}$ keV, and
optically thick  $\taue = 9.4\pm 1.3$ (for slab geometry). 
Reprocessed emission from the hard component is required in the spectrum,
with amplitude $\Or=0.5^{+0.6}_{-0.2}$ and
ionization parameter $\xi=(2.3^{+4.0}_{-2.1})\times 10^4$.
The relativistic smearing is not required by the data, but they are
consistent with smearing with $\Rin> 17\,\Rg$.
Setting the reflection amplitude $\Or=0$ and fitting all the other parameters 
of the continuum gives a bad fit with $\hi=95/26$.

Again, an alternative description of the soft spectral broadening can
be found using {\sc diskbb} with an additional blackbody ($\hi=16.1/22$).
However, the hybrid Comptonization does not provide a good fit.

\subsubsection{High State -- March 18th data}

Approximately two months after the peak of the outburst Nova Muscae 1991
made a transition to the High State, where its spectrum showed a
strong soft component with only very weak hard tail, if at all present 
(Ebisawa et al.\ 1994; Esin et al.\ 1997; see Fig.~\ref{fig:VHHS}). 
Ebisawa et al.\ (1994) noticed that the disc blackbody
description failed completely for these data ($\hi\sim 10$). 
We have modelled the spectrum obtained on 18th March. Applying the 
{\sc diskbb + thComp } model gives $\hi = 54/26$. %mar18_diskbb_thcompfe.xcm
The fit is bad, since the Comptonized component 
is steep, $\Gamma \approx 8$, so it makes a significant contribution to 
the soft component and,  as a consequence, misses the points at $E>10$ keV.
Thus again the soft component has to be described by a model broader
than the multi-colour disc blackbody. 

This can also be demonstrated by
ignoring data above 10 keV, thus leaving only the soft component 
(contribution from the hard component to $E<10$ keV is negligible)
and attempting to model it by a disc blackbody spectrum. 
The fit is very bad, $\hi=756/12$, even if $\NH$ is allowed to vary. 
The {\sc grad} models gives a better fit, $\hi=329/12$, but this is 
still far from satisfactory. These
data can be well fit by a Comptonized disc blackbody, giving $\hi=10.9/10$.
%mar18_thcompds_below10k.xcm

The overall  spectrum can be well fit using the {\sc compPS} model with a 
hybrid  electron energy distribution, $\hi = 18.0/24$.  
%mar18_compps_hybrid.xcm 
The low energy, thermal ($\kT\approx 3$ keV)
electrons Comptonize the seed, disc blackbody photons broadening somewhat 
the soft component, while the power law tail in the electron energy
distribution gives rise to the weak power law tail seen above 10 keV.
An additional blackbody also gives a good fit: {\sc wabs (diskbb + bbody
+ thComp)} gives $\hi=14.6/26$. % diskbb_bb_thcomp.xcm

% luminosity ?? $F=6\times 10^{-8}\,\ergs$

\subsection{Very High State of GRO~J1655-40}

GRO~J1655-40 went through a sequence of spectral/temporal states during
decline after the outburst observed in March -- September 1997. 
Analysis of the data by M\'{e}ndez et al.\ (1997) suggests that during the 
May 28 observation the source was in 
the VHS: the energy spectrum contained both a strong soft component 
and a power law tail. Likewise, the PSD appears to be a sum of a power law
component and a flat-topped component (see Takizawa et al.\ 1997). 

We use the 2.5--30 keV top layer {\it RXTE}/PCA and 
30--200 keV {\it RXTE}/HEXTE data for analysis.
The $\NH$ value towards this object is rather uncertain:
the $E(B-V)=1.15$ (Bailyn et al.\ 1995) corresponds to 
$\NHEBV=6.5\times\tentoto$ for standard extinction curve.
Fitting the {\em ASCA\/} data of the August 1995 observation  Zhang, 
Cui \& Chen (1997a) found $\NH=8.9\times \tentoto$, using a power law model
for the hard component, while Gierli\'{n}ski, Macio{\l}ek-Nied\'{z}wiecki 
\& Ebisawa  (2001) found
$\NH \approx 7\times \tentoto$ using a Comptonization model for the
hard component. Fitting the same data, we find $\NH$ values both higher
and lower than $6.5\times\tentoto$ depending on the model, 
therefore we let this parameter be free in our fits. 
We assume the source inclination is $i=70^{\circ}$.

We begin with PCA data alone. Unusually, the simple
model {\sc wabs (diskbb + thComp + rel-repr)} gives a good
fit, $\hi = 36.2/54$, % 14_top_diskbb_thcompds_freenh.xcm
with $\NH = (4.3\pm 1.4) \times\tentoto$.  Fixing $\NH$ at $\NHEBV$ gives 
$\hi=42.5/55$. % 14_top_diskbb_thcompds_nh0x65.xcm
Further broadening of the soft component is allowed by the data
although not required. Replacing the disc blackbody model with {\sc thComp}
yields $\hi = 33.0/53$, again for a cool
$\kT\approx 1$ keV, optically thick, $\taue\approx 10$ (for a slab
geometry) plasma, Comptonizing seed photons at a rather low temperature
$T_0 \approx 0.3$.
%The same results are obtained with the {\sc thComp} model. 
These results are very similar to those of Zhang et al.\ (2000), who
fitted similar models to the {\em ASCA\/} data of GRO~J1655-40 and
GRS~1915+105. Comptonization of the soft component in this source,
in high luminosity states, is also reported by Kubota, Makishima \& Ebisawa 
(2001).

\begin{figure}
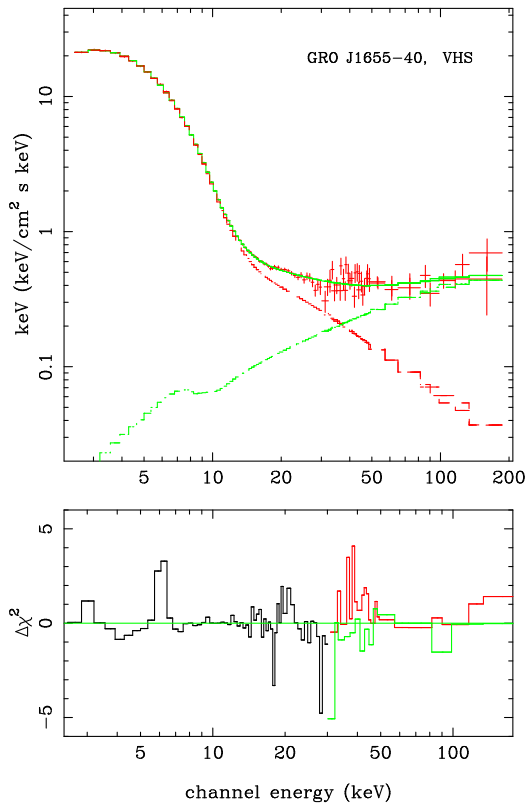

 \epsfxsize = 0.45\textwidth
 \epsfbox[0 200 600 720]{pca_hexte_fit.ps}
 \epsfxsize = 0.45\textwidth
 \epsfbox[0 400 600 700]{pca_hexte_chi.ps}
 \caption{Model fit to the VHS {\it RXTE\/} data of GRO~J1655-40
($\hi=90/113$). 
The model comprises the hybrid Comptonization and additional thermal 
Comptonization. The reprocessed component is clearly detected,
with amplitude $\Omega/(2\pi) \approx 1$, high ionization, 
$\xi \approx 10^5$ and relativistic smearing, $\Rin \approx 16\Rg$.
\label{fig:1655spec}}
\end{figure}

Again, the alternative descriptions of the broad soft excess give
excellent fits to the data. A hybrid electron distribution
gives $\hi=33.2/55$, while an                   % 14_top_compps_hybrid.xcm 
additional blackbody gives $\hi = 35.2/53$,     %14_top_diskbb_bb_thcompfe.xcm
both fits with $\NH = \NHEBV$.

The central black hole in this object is postulated to be a rotating
one based on the observed high temperature and small amplitude of the
thermal disc emission (Zhang et al.\ 1997a).  Gierli\'{n}ski et al.\
(2001) infer black hole spin $a=0.7-0.8$, fitting a relativistic disc
spectrum model, with photon propagation in the Kerr metric to the 1995
{\it ASCA\/} data. However, in order to obtain a good fit,
Gierli\'{n}ski et al.\ need to introduce a smeared edge at 2--3 keV,
to model distortions present in the spectrum.  Replacing the disc
blackbody model in our fits with a disc blackbody in Kerr metric we
obtain $\hi= 44.6/54$, % top_kerrd_thcompfe_rin_free.xcm for $\NH$
fixed at $\NHEBV$, but with the inner disc radius and central mass as
free parameters (to compensate for the fixed value of the black hole
spin in our model; see discussion in Gierli\'{n}ski et al.\
2001). Thus using the Kerr metric disc model does not improve the fit
compared to the simple disc blackbody model. In view of the complex
soft component shape which is generally seen in the soft spectral
states we caution strongly against interpreting any broadening of
the thermal emission solely as relativistic effects (although such
effects should be present).

The above models are no longer adequate when the HEXTE data are added.
An additional, harder continuum component is required to model the broadband 
spectrum, as already found for the VHS spectra of XTE~J1550-564 by 
Wilson \& Done (2001).
A possible description of the spectrum is provided by the hybrid
Comptonization model with an additional, thermal Comptonization.
The slopes of the two Comptonized tails (the non-thermal tail and
the additional Comptonization) are not well constrained individually,
so we fit the model fixing the slope of the non-thermal electron distribution
at $\gamma=5$ (i.e.\ steeper than in the fit to PCA data alone). 
We allowed both continua to give rise to the reprocessed component,
but with the same parameters ($\Omega/(2\pi)$, $\xi$ and $\Rin$).
The overall model,
{\sc wabs(compPS$_{\rm hybrid}$ + rel-refl + thComp + rel-repr)},
gives a good fit to the data, $\hi=90/113$ (Fig.~\ref{fig:1655spec}). 
The reprocessed component is clearly detected in the spectrum:
without it the fit is much worse, $\hi=158/116$. 
% 14_pca_hexte_compps_hybrid_thcomp_no_refl.xcm
The reflection amplitude $\Omega/(2\pi) \approx 1$  
(lower limit $\approx 0.5$ but with poorly constrained upper limit), 
the ionization parameter  $xi\approx 10^5$, and the inner disc radius 
$\Rin 16\,\Rg$. 
This differs from the results of Gierli\'{n}ski et al.\ (2001), 
who did not find any significant reflection features in {\it ASCA\/} data 
of this source.

\section{Discussion}

We have re-analysed archival data of a number of SXT sources in soft
spectral states in an attempt to characterize their spectral properties
in more detail than has generally been done so far. In most previous analyses
of IS, HS and VHS spectra the assumed spectral model consisted of a disc
blackbody 
emission, a power law, a smeared edge and a gaussian line near 5--9 keV.
Of these, the edge and the line are the least accurate representation
of the underlying physical processes. We replace them with a more 
physical model of X-ray reprocessing by an optically thick plasma. Also,
a power law is an inaccurate approximation to Comptonized emission
at energies close to the seed photons as true 
Comptonized spectra have a low energy cutoff. In the soft spectral
states the seed photons from the disc are within the observed
bandpass, so such effects become important.

Two results from the above data analyses seem to be robust: firstly, the soft
component {\it cannot\/} be adequately described by the simple disc
blackbody model and, secondly, the reprocessed component is present in
the data, it is strongly ionized and smeared.

The complexity of the soft component is not merely an artifact of
using the approximate {\sc diskbb} model. Including the torque-free 
boundary condition, colour temperature and
relativistic effects in photon propagation does not
substantially change the overall shape of the spectrum, although the derived
model parameters can be significantly different (e.g.\ Ebisawa 1991; 
Gierli\'{n}ski et al.\ 1999; Merloni, Fabian \& Ross 1999). 
Direct fitting of our relativistic disc model {\sc grad}, which includes 
the corrections, supports our claim. Although the best fits are 
somewhat better than the model employing {\sc diskbb},
they are much worse than fits employing Comptonized black body. Similarly,
using the Kerr metric disc emission model did not provide better description
of the spectrum of GRO~J1655-40 compared to the simple disc blackbody
model. The soft component is substantially broader than optically thick disc
models, irrespective of relativistic corrections.

Using the \ginga or \xte data we are not able to differentiate between
various possible modifications of the disc blackbody spectrum which
ensure a good description of data. The two main possibilities are
additional Comptonization of the disc emission or 
an additional blackbody component. We discuss each of these in turn.

\subsection{The complex soft component: additional blackbody}

One possibility to explain the complex shape of the soft component is
to assume that it is a sum of a disc blackbody and additional
blackbody emission, indicating that some fraction of the disc material
is heated to rather higher temperatures.  Physically we might expect
this if the hard X--ray power law arises from magnetic
reconnection. The regions of the disc below the intense flares are
then heated by the irradiating flux and the thermalized
(non--reflected) radiation would be emitted as a quasi--blackbody
spectrum at temperatures somewhat higher than the mean disc emission
(Haardt, Maraschi \& Ghisellini  1994).
Alternatively, for a central source geometry, the inner disc is additionally
heated by irradiation. In all these
models the additional blackbody is powered ultimately by the hard
component.

Energetically this is certainly feasible for the IS spectra. 
Consider, for example, the GS~1124-68 May 18th data.
The bolometric (0.01--1000 keV) flux from this model is (in units of
$10^{-9}\ \ergs$):
$\Ftot \approx 25$, the hard X--ray flux is $\Fhard \approx 2.0$, while
the flux in the additional blackbody is $\Fbb = 0.88 \approx 0.44\Fhard$. 
The reprocessor albedo is observed to be $a\approx 0.75$
(ionization of $\xi\approx 3000$, so that Fe{\sc XXV} and 
Fe{\sc XXVI} are the dominant ionization stages of iron).
Thus the thermalized fraction is expected to be
$1-a \approx 0.25$. Given the uncertainties in the modeling this is
probably consistent with the observed fraction $\approx 0.4$.

The observed flux $\Fbb \approx 9\times 10^{-10}\, \ergs$ at $T_{\rm
bb} \approx 0.7$ keV implies an emitting area $S=4\pi d^2 \Fbb/(\sigma
T^4 \cos i)$, where $\sigma$ is the Stefan-Boltzmann constant. This
gives a tiny area, $S\approx 10^{13}\,{\rm cm^2} \approx 13\,\Rg^2$,
assuming distance $d=3.5$ kpc, $M=6\,\MSun$ and neglecting the colour
temperature correction. This is consistent with magnetic flare
illumination, if the illuminating area is small, i.e.\ if the flares
are located not too high above the disc (Nayakshin \& Kazanas 2001).
In the central source geometry the distribution of irradiation flux is
somewhat steeper than $r^{-3}$ close to the inner disc radius (see fig.~A2
in \.{Z}ycki et al.\ 1999a). This gives an excess of emission,
if modelled by the {\sc diskbb} model, but the excess is not sufficient
to account for the observed shape of the soft component. Partial
overlap between the central source and the disc might solve the problem.

However, the soft component is overwhelmingly dominant in the HS
spectra  and reprocessing of the very small hard X--ray flux is not
consistent with the flux required for the additional blackbody.
An extreme example here is the HS GS~1124-68 March 18th dataset, 
where the hard
X--ray ($E>10$ keV) flux is less than 1 per cent of the total flux,
while the additional blackbody component contains $\sim 1/3$ of the total flux.
The situation is not so extreme for the VHS spectra, where there is
some hard X--ray emission. The observed fluxes of hard X--rays
are rather smaller than the fluxes of the additional blackbody components, 
which rules out the possibility of the blackbody being the thermalized
hard X--ray illumination, if isotropic emission of the latter is assumed. 
The peak spectrum of GS~1124-68 (Jan 16th; Sec~\ref{sec:jan16}) is a good 
example here: the fluxes in the 0.01--$10^3$ keV band (in units of 
$10^{-8}\ {\rm ergs}\ {\rm cm}^{-2}\ {\rm s}^{-1}$) are: $\Ftot = 28.3$,
$\Fdbb = 17.1$, $\Fhard = 3.4$ and $\Fbb = 7.8 > \Fhard$. 
If the hard X--rays were beamed towards the disc, both the thermalized 
emission and the reprocessed component would be enhanced. The amplitude
$\Or$ is not always well constained in VHS data due to complex continuum
curvature; for example for the Jan 16th data $\Or=1.8^{+0.6}_{-1.0}$,
while for the Feb 13th $\Or = 0.5^{+0.6}_{-0.2}$.

\subsection{The complex soft component: Comptonization}

The two-component description of the disc emission does not seem appropriate
for the brighter spectra, especially those lacking a strong hard
component. 
The additional Comptonization of the disc emission provides an alternative
description to those (and all other) spectra. 
The parameters of the Comptonizing plasma are not well
constrained, but generally a
cool, $\kT = 2-4$ keV, optically thick, $\taue \sim 5$, plasma gives
the best fit. The main difference between this and the same spectral index 
Comptonization from a hotter, optically thinner plasma 
is in the  possible contribution of the seed photons
to the final spectrum for $\taue < 1$. However, this subtle difference
is not sufficient to uniquely select the best fit solution, given the rather 
poor energy resolution and low energy bandpass of the data used here. 

We can envision a number possibilities for the origin of the Comptonization
of disc emission. 

\subsubsection{Warm atmosphere of the accretion disc}

The soft photons from the disc interior can be
Comptonized as they escape through a hot, ionized topmost layer
of the disc. 

If the Comptonizing layer is heated by external irradiation only,
i.e.\ viscous heating is negligible, its maximum temperature is the
inverse Compton temperature, $\Tic$, at which Compton heating is balanced by
Compton cooling. This temperature can be estimated from
\begin{equation}
4 k \Tic \int \uE\,dE = \int E \uE  \sigmaKN(E) \,dE,
\end{equation}
% \left(1-{21 E \over 5 \me c^2}\right)
where $\uE$ is the radiation energy density, and $\sigmaKN(E)$ is
the  Klein--Nishina cross section. For a spectrum consisting of a strong 
soft component and
a steep power law, $\Tic$ is not much larger than the temperature of the 
soft component. For example, for the May 18th data of GS~1124-68 we calculate
$\Tic \approx 0.9$ keV, and similar values are 
obtained for the other IS spectra. This is  lower than the inferred
temperature of the Comptonizing plasmas, implying that there {\it
must\/} be additional heating
of the plasma through e.g.\ viscous dissipation (Janiuk, Czerny \& Madejski, 
2001).
Only for the 11th Jan spectrum of GS~1124-68 is $\Tic\approx 4$ keV
comparable to the required $\kT$ of the Comptonizing plasma.

This picture then  implies a three-phased accretion scenario:
the hot, $\kT\sim 100$ keV, plasma with the Comptonization parameter 
$y\sim 1$, producing the hard X--ray continuum; the warm, 
$\kT = 1$--10 keV plasma with $y\sim 0.3$, responsible for 
the additional Comptonization of the disc emission; and the optically thick,
'cold', $\kT \sim 0.1$--1 keV disc. 
A similar accretion scenario was also suggested for the Seyfert 1
galaxy NGC~5548 (Magdziarz et al.\ 1998). The clear soft X-ray excess
component in that object could be well modelled by a Comptonization
of accretion disc photons in a warm ($\kT \sim 300$ eV), optically thick
($\taue \sim 30$) plasma. Magdziarz et al.\ speculated that the plasma 
could form a transition region between the outer, cold accretion disc
and the inner hot flow. Similar Comptonized soft X--ray excesses are
present in two Narrow Line Seyfert 1 galaxies: PG~1211-143
(Janiuk et al.\ 2001), and PKS~0558-504 (O'Brien et al.\ 2001).
A possible analogy of such three-phased
configuration around accreting black holes with the Solar corona was
suggested by Zhang et al.\ (2000). 

However, the warm plasma cannot form a continuous later covering the
cold disc. It is optically thick and completely ionized
so would produce no spectral features, contrary to results of our
data analyses. More robust (given the uncertainties in optical depth 
and temperature) arguments against such a geometry are provided by 
considering the energy balance between the cool disc and warm skin. 
Following the approach of Haardt \& Maraschi (1991), the energy
balance can be written as

\begin{equation}
\label{equ:fseed}
1-f + \eta f = \Fseed,
\end{equation}
while for the warm skin
\begin{equation}
\label{equ:fcomp}
A \Fseed = \Fcompt = f,
\end{equation}
where $f$ the fraction of energy 
dissipated in the warm skin (disc dissipation $1-f$), $\eta$ is
the fraction of Comptonized emission returning to the cold disc,
$A$ is the Comptonization amplification factor, $A\equiv \Fcompt/\Fseed$,
and all the fluxes are normalized to the total gravitational energy dissipated
in the two layers. Solving Eq.~(\ref{equ:fseed}) and Eq.~(\ref{equ:fcomp}),
we obtain the condition for the existence of solution,
\begin{equation}
 \eta A \le 1.
\end{equation}
Curiously, the Comptonization parameters from the fits to GRO~J1655-40 imply 
$A\approx 1.24$ and $\eta\approx 0.82$, thus $|\eta A-1| < 0.02$. This,
in turn, implies $f$ very close to 1, i.e.\ the entire energy dissipation in 
the warm skin, with the cold disc merely reprocessing and thermalizing
the intercepted radiation.

The situation is analogous to that with a continuous hot corona overlapping
an optically thick disc (Haardt \& Maraschi 1991). Even in the extreme case
of $f=1$ the produced spectra are not harder than $\Gamma\approx 2$,
because the geometry alone implies $A \le 2$.
Any harder spectra imply e.g.\ a patchy corona, coronal outflow, etc. 
(e.g.\ Stern et al.\ 1995; Beloborodov 1999).

A possible solution to the reprocessed features problem 
is to assume that the scattering layer appears
only in the inner part of the disc, close to the radius of maximum
effective temperature. The reprocessing features could then be
produced further out, where the temperature is lower enabling iron to
recombine to H- and He-like ions, as observed. The relativistic
smearing of the iron features, usually corresponding to $\Rin \ge
10\,\Rg$ supports this possibility.

\subsubsection{Hybrid electron energy distribution in the Comptonizing plasma}

Another possibility for the Comptonization of the soft photons is to
assume a hybrid, thermal/non-thermal distribution of electron energies
in an accretion disc corona
(see Coppi 1999 for extensive discussion of relevant physical processes). 
The harder power law spectrum extending above $\sim 10$
keV would then be a result of single inverse Compton upscattering of the
disc photons by energetic electrons from the power law tail of the
distribution. The Comptonized tail of the soft component is a result of 
scattering of the photons by the low energy electrons, with roughly
Maxwellian distribution.  Such model was successfully
applied to the soft state data of Cyg X-1 (Gierli\'{n}ski et al.\ 1999), where
the soft component shows a clear excess above a disc blackbody component,
while the hard power law extends without a break to at least 1 MeV
(Grove et al.\ 1998). 
Fitting the {\sc CompPS} Comptonization model (Poutanen \& Svensson 1996)
with such hybrid electron distribution we found that indeed it provides
a good description of some of the spectra: the IS spectra of all three
analysed objects are well described by it, as well as some VHS spectra:
the peak spectrum of GS~1124-68 and the spectrum of GRO~J1655-40.
On the other hand, we were not able to find an acceptable description of
the Jan 11 and Feb 13 of GS~1124-68 with this model.

\subsubsection{Weak thermal corona}

Yet another possibility for the Comptonization of soft photons is a weak, 
accreting corona, similar to that studied by Witt, Czerny \& \.{Z}ycki (1997),
and Janiuk, \.{Z}ycki \& Czerny (2000). In these models the disc--corona 
transition is governed by the condition of thermal equilibrium between 
the two phases of accreting plasma. This results in a rather weak
corona, dissipating up to $\sim 40$ per cent of gravitational energy. Details
of the solution depend upon the role of advective energy transfer (compare
solutions in Janiuk \& Czerny 2000 and Janiuk et al.\ (2000), 
but generally
the corona is optically thin, $\taue \le 0.15$ and its temperature varies 
between $\approx 70$ keV at $R \approx 10\,\Rg$ to 150 keV further away.
The resulting spectra are rather steep and roughly match the
Comptonized excesses in the IS spectra, but such a weak corona would not
be able to account for the strong Comptonization observed in e.g.\ Jan 16th
spectrum of GS~1124-68.
The model cannot account for the origin of the hard component seen in IS and 
VHS. There is, however, growing evidence of a very close link between hard
X--ray and radio emission (see Fender 2000, 2001 for reviews). If the two
are indeed linked, one may envision the hard X--rays originating from
e.g.\ a base of a jet.

\subsection{X--ray reprocessing}
\label{sec:repro}

Spectral features near 6--9 keV are commonly observed in spectra of
accreting compact objects, and they are indeed present in the spectra analysed
in this paper. Reprocessing  of hard X--rays  by an optically 
thick plasma (Compton reflection, photo-absorption, fluorescent and 
recombination emission) is
the standard interpretation of these features. It has been tested on numerous 
spectra of all kind of accreting objects, although most of the modeling work 
has been done with low/hard state spectra. The problem with soft states spectra
is the complicated shape of the continuum emission near 5--10 keV, where
the soft and hard component contribute roughly equally. In our work we do see 
that the parameters of the reprocessed component in the soft states spectra
are dependent on the
continuum model adopted. However, two results are robust from the work 
done so far:
\begin{itemize}
 \item the very presence of the reprocessed component, i.e.\ we have not 
  been able to find a continuum description that would make the reprocessed 
  component unnecessary,
 \item high level of ionization of the reprocessor: again, irrespectively
  of the continuum model, the reprocessing features correspond to highly
  ionized plasma, where most abundant Fe ions are more strongly ionized than
 Fe{\sc XVI} (M-shell
 electrons removed), and in most cases the He- and H-like ions.
\end{itemize}
The third result is that the spectral features are further broadened and 
smeared compared to predictions of the simple model. We model the smearing
as the relativistic Doppler effect and gravitational redshift, but it has to 
be kept in mind that Comptonization in the reprocessor is quite significant,
especially for strong ionization
(e.g.\ Ross et al.\ 1999).
Therefore the quantitative determinations
of the inner radius of the reprocessing accretion disc may be overestimated.

A potential alternative to the reprocessing scenario might be 
absorption/emission
in winds which are almost certainly launched from the vicinity of
central objects accreting at $\mdot$ close to the Eddington limit
(e.g.\ Proga 2000 and references therein). The characteristic 'P-Cygni'
emission/absorption profiles might turn out to be compatible with the
data.
This idea has not been explored in any details and its quantitative tests are 
not possible at present, as the spectral predictions of the models are not 
ready yet to be quantitatively confronted with the data.

\begin{figure}
 \epsfxsize = 0.45\textwidth
 \epsfbox[0 300 600 700]{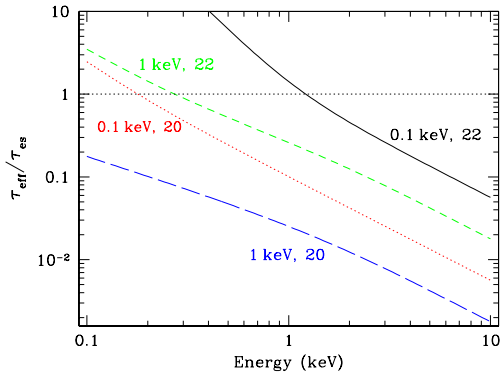}
 \caption{Effective absorption cross-section (free-free process only) as 
 a function of energy, for a number of values of the gas temperature, $T$,
 and H number density, $\nH$. Values of $T$ and $\log(\nH)$ are plotted
 next to corresponding curves. Since the ratio 
 $\tau_{\rm eff}/\tau_{\rm es} < 1$ for $E>1$ keV, the incident X--rays
 will not be thermalized in the upper disc layers, where the process
 of Compton reflection is taking place.
\label{fig:ffkapeff}}
\end{figure}

One potential problem that might affect our conclusions is that the 
computations of the reflected continuum used in the {\sc rel-repr} model 
(Magdziarz \& Zdziarski 1995) assume that 
the only X--ray absorption process is the photo-electric absorption.
However, the density in accretion discs around stellar mass black holes 
is high enough for the free-free process to become important. Another
way of formulating the problem is to notice that -- by the very definition
of a soft state -- the X--rays at $E\sim 0.5$ keV are thermalized.
If so, then the reflected continuum could have a rather different shape
at low energies than that used in our model. Precise computations of
the reflected continuum in such conditions are not possible at present,
as no photo-ionization code can handle densities 
$\sim 10^{20}\ {\rm cm^{-3}}$ and thermalization of incident X--rays.
We can however estimate whether the incident X--rays
at $E \ge 1$ keV are likely to be thermalized or not. 
Solving the vertical structure
of an $\alpha P$ disc around $10\,\MSun$ black hole at $R=10\,\Rg$ 
(see e.g.\ R\'{o}\.{z}a\'{n}ska et al.\ 1999) we find the H number density
at the disc equator $\nH \sim 10^{22}\,{\rm cm^{-3}}$. We compute the 
free--free absorption 
cross-section (Rybicki \& Lightman 1979) as a function of energy for a number
of values of gas temperature and density. The results, plotted as 
the effective absorption, $\kappa_{\rm eff} = \sqrt{\kappaabs (\kappaabs + 
 \kappaes)}$ in units of the Thomson cross section as a function of energy, 
are shown in Fig.~\ref{fig:ffkapeff}.
Since the effective absorption coefficient is smaller than 1 for $E>1$ keV,
the incident X--rays will not be thermalized in the uppermost 
$\tau_{\rm es} \sim 1$ layer, where Compton reflection takes place.
Thus the adopted 
model for the Compton reflected continuum is appropriate.

\subsection{Disc stability}

Perhaps the most puzzling behaviour of SXTs, in the context of the 
$\alpha$-discs model (Shakura \& Sunyaev 1973), is the strong soft thermal 
emission observed when the sources radiate at a significant fraction
of the Eddington luminosity, when the $\alpha P_{\rm tot}$-discs
are unstable (Kato et al.\ 1998), and so
perhaps they should not exist. On the other hand, these discs are stable
for low $L/\LEdd$ (LS) when, observationally, they seem not to be present
in the central regions of GBH. While the latter problem seems to have
found its solution in the mechanism of disc evaporation (R\'{o}\.{z}a\'{n}ska
\& Czerny 2000 and references therein), the former one is far from
any satisfactory explanation.

\begin{figure}
 \hfil\epsfxsize = 0.45\textwidth
 \epsfbox[20 200 600 700]{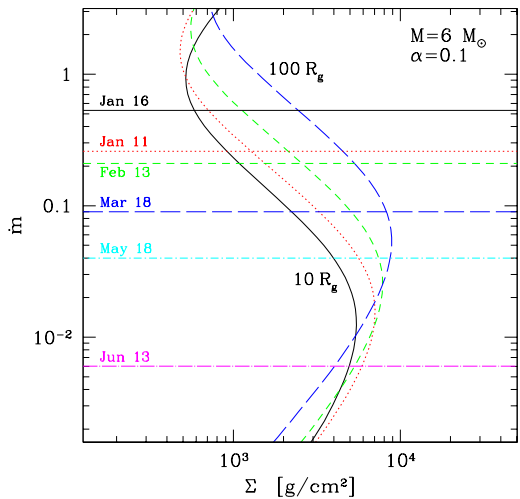}\hfil
\caption{The surface density -- mass accretion rate diagrams for an
$\alpha\Ptot$ accretion disc around a $6\,\MSun$ black hole. The four thick
curves are computed for $R=10,\ 30,\ 50\ {\rm and}\ 100\,\Rg$. The parts
of the curves with negative slope represent unstable solutions. The thin 
horizontal lines show estimated values of ${\dot m} \equiv {\dot M}/\MEdd$ 
for GS~1124-68, assuming $d=3.5$ kpc. 
\label{fig:smdot}}
\end{figure}

Here we construct the $\dot m$--$\Sigma$
(accretion rate -- surface density) diagram and mark the position
of GS~1124-68, to quantitatively demonstrate the 
extent of the instability. This extends studies done by Gierli\'{n}ski
et al.\ (1999), who argued that in the soft state of Cyg~X-1 the accretion 
disc could remain on the stable branch, down to the last stable orbit.
However, we note that the soft state of Cyg X-1 seems to correspond
to an Intermediate State of GS~2000+25 or GS~1124-68 rather than a true 
High State, since 
its bolometric luminosity is rather low, only $\approx 0.03\, \LEdd$ 
(Gierli\'{n}ski et al.\ 1999), not much higher than a typical low state 
luminosity (Zhang et al.\ 1997b; Di Salvo et al.\ 2001).
We estimated the bolometric luminosities of GS~1124-68  from models using
either disc blackbody or Comptonized disc blackbody spectra, which should give
an upper limit to the bolometric correction. We further assumed a $6\,\MSun$
black hole at $d=3.5$ kpc. Since the largest value of $d$ found in
literature is $\approx 5$ kpc (Esin et al.\ 1997), the uncertainty
of $d^2$ should be less than factor of 2.
Figure~\ref{fig:smdot} shows positions of the source on the $\dot m$--$\Sigma$
diagram, demonstrating that the disc was stable in LS, but it was unstable
(at least in some inner region) whenever the source was observed in a soft 
state. One might then speculate that the result of the instability is
a slight reconfiguration of the disc, so that the disc emission is still
optically thick, but with the additional Comptonization in some warm plasma.
There is a number of instabilities operating in radiation pressure dominated
discs (Kato et al.\ 1998; Gammie 1998; Blaes \& Socrates 2000), 
and it is possible that
their outcome may be a clumpy accretion flow (Krolik 1998). On the other
hand, the amplitude of the soft component sees to be constant during
the part of the decline when the spectral state is soft (Ebisawa 1991),
indicating that the covering factor of the clumps cannot to be low, so 
that the disc effectively looks continuous.

The highest luminosity spectra seem to need rather different parameters of 
the Comptonization compared to the IS spectra.
Perhaps, the disc structure changes when  the disc solution
is located on the upper stable branch (so called 'slim discs': optically 
thick, with advective cooling; Abramowicz et al.\
1988). Alternatively, if the viscous heating scales with the gas
pressure rather than the total (gas plus radiation) pressure then the
instability is avoided (Stella \& Rosner 1984). While such a scaling
might be expected for an MHD dynamo origin of the viscosity (Stella \& 
Rosner 1984), this has not yet been confirmed by numerical simulations.
An interesting alternative viscosity prescription is implied by hydrodynamic
experiments on rotating flows (Richard \& Zahn 1999). For a Keplerian disc,
the viscosity can be written as $\nu = \beta \OmegaK r^2$, where $r$ is the
distance from the rotation axis and the coefficient $\beta \sim 1/{\cal R}e$,
where ${\cal R}e$ is the Reynolds number. This prescription gives 
disc solutions that are both thermally and viscously stable 
(Hur\'{e}, Richard \& Zahn 2001).

\section{Conclusions}

We model the spectra of Soft X--ray Transients in soft states 
using more physical models for the emission. We use reflection models as
opposed to phenomenological broad line/smeared edge features, and
Comptonization models which include the low energy cutoff at
the seed photons as opposed to a continuous power law.
We find that the strong soft component cannot be described by simple
disc models (disc blackbody with colour temperature and/or
relativistic corrections). There is additional broadening of the soft
emission, indicating the presence of intermediate (between the cool main disc
and the hot X--ray plasma) temperature material. This can be modelled
either as an additional blackbody, or as additional Comptonization. 
Both could be plausibly present. Magnetic reconnection above the disc could
produce an X--ray heated spot underneath the
flare, and/or the electrons in the flare itself may have a hybrid
distribution, with a thermal tail at low energies which
additionally Comptonizes the disc emission as well as a non--thermal
power law to give the hard X--ray flux. 
Alternatively (additionally?) there could be a warm skin over
the inner disc which Comptonizes the highest temperature emission from 
the disc. 

An ionized reflected component is {\it always\/} detected irrespective
of the detailed model used to describe the soft component.

\section*{Acknowledgments} 
 
This work  was partly supported by grant no.\ 2P03D01718  of the Polish 
State Committee for Scientific Research (KBN) and by British-Polish 
Joint Research Collaboration Programme.

{}


\begin{thebibliography}{}

 \bibitem[]{}
   Abramowicz M. A., Czerny B.,  Lasota J. P., Szuszkiewicz E., 1988,
      ApJ, 332, 646
 \bibitem[]{}
   Arnaud K. A., 1996, in Jacoby G., Barnes J., eds, ASP Conf. Ser. Vol. 101,
       Astronomical Data Analysis Software and Systems V. Astron. Soc. Pac.,
       San Francisco, p. 17
 \bibitem[]{}
   Bailyn C. D. et al., 1995, Nature, 374, 701
 \bibitem[]{}
   Belloni T., van der Klis M., Lewin W. H. G., van Paradijs J., Dotani T.,
      Mitsuda K., 1997, A\&A, 322, 857
 \bibitem[]{}
   Beloborodov A. M., 1998, MNRAS, 297, 739
 \bibitem[]{}
   Beloborodov A. M., 1999, ApJ, 510, L123
 \bibitem[]{}
   Blaes O., Socrates A., 2000, ApJ, submitted, (astro-ph/0011097)
 \bibitem[]{}
   Burderi L., King A. R., Szuszkiewicz E., 1998, ApJ, 509, 85
 \bibitem[]{}
   Callahan P. J., Garcia M. R., Filippenko A. V., McLean I., Teplitz H.,
    1996, ApJ, 470, L57
 \bibitem[]{}
   Cannizzo J. K., 1993, in Wheeler J. C., ed, Accretion in Compact Stellar 
     Systems. World Scientific, Singapore, p. 6
 \bibitem[]{}
   Chen W., Shrader C. R., Livio M., 1997, ApJ, 491, 312
 \bibitem[]{}
   Coppi P., 1999, in Poutanen J., Svensson R., eds, ASP Conf. Ser. Vol. 161,
    High Energy Processes in Accreting Black Holes. Astron. Soc. Pac.,
    San Francisco, p. 375
 \bibitem[]{}
   Di Salvo T., Done C., \.{Z}ycki P. T., Burderi L., Robba N. R., 2001,
    ApJ, 547, 1024
 \bibitem[]{}
   Done C., 2001, AdSpR, in press, (astro-ph/0012380)
 \bibitem[]{}
   Done C., Mulchaey J. S., Mushotzky R. F., Arnaud K. A.,  1992, ApJ, 395, 275
 \bibitem []{}
   Ebisawa K., 1991, PhD thesis, Univ.\ of Tokyo
 \bibitem[]{}
   Ebisawa K. et al., 1994, PASJ, 46, 375
 \bibitem[]{}
   Esin A. A., McClintock J. E., Narayan R., 1997, ApJ, 489, 865
 \bibitem[]{}
   Fabian A. C., Rees M. J., Stella L., White N. E., 1989, MNRAS, 238, 729
 \bibitem[]{}
   Fender R., 2001, in Kaper L., van den Heuvel E. P. J., Woudt P. A., eds,
     Black Holes in binaries and galactic nuclei. Springer-Verlag, p. 193,
     (astro-ph/9911176)
 \bibitem[]{}
   Fender R., 2000, Ap\&SS, in press, (astro-ph/0010613)
 \bibitem[]{}
   Gammie C. F., 1998, MNRAS, 297, 929
 \bibitem[]{}
   George I. M., Fabian A. C., 1991, MNRAS, 249, 352
 \bibitem[]{}
   Gierli\'{n}ski M.,  Zdziarski A. A., Done C., Johnson W. N.,  Ebisawa K.,
    Ueda Y., Phlips F., 1997, MNRAS, 288, 958
 \bibitem[]{}
   Gierli\'{n}ski M.,  Zdziarski A. A., Poutanen J., Coppi P., Ebisawa K.,
    Johnson W. N., 1999, MNRAS,  309, 496
 \bibitem[]{}
   Gierli\'{n}ski M.,  Macio{\l}ek-Nied\'{z}wiecki A., Ebisawa K., 2001, MNRAS,
      in press (astro-ph/0103362)
 \bibitem[]{}
   Gilfanov M., Churazov E., Revnivtsev M., 2000,  in Proc.\ of the 5th 
   CAS/MPG  Workshop on High Energy Astrophysics, in press, (astro-ph/0002415)
 \bibitem[]{}
   Grove J. E., Johnson W. N., Kroeger R. A., McNaron-Brown K. Skibo J. G.,
    Phlips B. F., 1998, ApJ, 500, 899
 \bibitem[]{}
   Haardt F.,  Maraschi L., 1991, ApJ, 380, L51
 \bibitem[]{}
   Haardt F.,  Maraschi L., Ghisellini G., 1994, ApJ, 432, L95
 \bibitem[]{}
   Hayashida K., Inoue H., Koyama K., Awaki H., Takano S., 1989, PASJ, 41, 373
 \bibitem[]{}
   Homan J., Wijnands R., van der Klis M., Belloni T., van Paradijs J., 
      Klein-Wolt M., Fender R., M\'{e}endez M., 2001, ApJS, 132, 377
 \bibitem[]{} 
   Hur\'{e} J.-M., Richard D., Zahn J.-P., 2001, A\&A, 367, 1087
 \bibitem[]{} 
   Janiuk A., Czerny B., 2000, NewA, 5, 7
 \bibitem[]{} 
   Janiuk A., Czerny B., Madejski G. M., 2001, ApJ, in press 
 \bibitem[]{} 
   Janiuk A., \.{Z}ycki P. T., Czerny B.,  2000, MNRAS, 314, 364
 \bibitem[]{} 
   Kato S., Fukue J., Mineshige S., 1998, Black Hole Accretion Disks.
         Kyoto University Press, Kyoto 
 \bibitem[]{} 
   King A. R., Ritter H., 1998, MNRAS, 293, L42 
 \bibitem[]{}
   Krolik J. H., 1998, ApJ, 498, L13
 \bibitem[]{}
   Kubota A., Makishima K., Ebisawa K., 2001, ApJ, submitted, 
                                      (astro-ph/0105426)
 \bibitem[]{}
   Laor A., 1991, ApJ, 376, 90
 \bibitem[]{}
   Lightman A. P., White T. R., 1988, ApJ, 335, 57
 \bibitem[]{}
   Matt G., Perola G. C., Piro L., 1991, A\&A, 247, 25
 \bibitem[]{}
   Magdziarz P., Zdziarski A. A., 1995, MNRAS, 273, 837
 \bibitem[]{}
   Magdziarz P.,  Blaes O. M., Zdziarski A. A., Johnson W. N., Smith D. A., 
      1998, MNRAS, 301, 179
 \bibitem[]{}
   M\'{e}ndez M., Belloni T., van der Klis M., 1998, ApJ, 499, L187
 \bibitem[]{}
   Merloni A., Fabian A. C., Ross R. R.,  1999, MNRAS, 313, 193
 \bibitem[]{}
   Meyer F., Meyer-Hofmeister E., 1994, A\&A, 288, 175
 \bibitem[]{}
   Miller J. M., Fox D. W., Di Matteo T., Wijnands R., Belloni T., Pooley D.,
   Kouveliotou C., Lewin W. H. G. 2001, ApJ, 546, 1055
 \bibitem[]{}
   Miyamoto S., Kitamoto S., Hayashida K., Egoshi W., 1995, ApJ, 442, L13
 \bibitem[]{}
   Mitsuda K. et al., 1984, PASJ, 36, 741
 \bibitem[]{}
   Morrison R., McCammon D., 1983, ApJ, 270, 119
 \bibitem[]{} 
   Nayakshin S., Kazanas D., 2001, ApJ, 553, L141
 \bibitem[]{} 
   Nayakshin S., Kazanas D., Kallman T. R., 2000, ApJ, 537, 833
 \bibitem[]{}
   O'Brien P. T. et al., 2001, A\&A, 365, L122
 \bibitem[]{}
   Osaki Y., 1996, PASP, 108, 39
 \bibitem[]{}
   Poutanen J., Svensson R., 1996, ApJ, 470, 249
 \bibitem[]{}
   Proga D., 2000, ApJ, 538, 684
 \bibitem[]{}
   Richard  D., Zahn J.-P., 1999, A\&A, 347, 734
 \bibitem[]{}
   Ross R. R., Fabian A. C., Brandt W. N., 1996, MNRAS, 278, 1082
 \bibitem[]{}
   Ross R. R., Fabian A. C., Young A. J., 1999, MNRAS, 306, 461
 \bibitem[]{}
  R\'{o}\.{z}a\'{n}ska A., Czerny B., 1996, Acta Astron., 46, 233
 \bibitem[]{}
  R\'{o}\.{z}a\'{n}ska A., Czerny B., 2000, A\&A, 360, 1170
 \bibitem[]{}
  R\'{o}\.{z}a\'{n}ska A., Czerny B., \.{Z}ycki P.T., Pojma\'{n}ski G.,
      1999, MNRAS,  305, 481
 \bibitem[]{}
  Rutledge R. E. et al., 1999, ApJS, 124, 265
 \bibitem[]{}
   Rybicki G. B., Lightman A. P. 1979, Radiative Processes in Astrophysics.
    John Wiley \& Sons, New York
 \bibitem[]{}
   S\'{a}nchez-Fern\'{a}ndez et al., 1999,  A\&A, 348, L9
 \bibitem[]{}
   Shakura N. I., Sunyaev R. A., 1973, A\&A, 24, 337
 \bibitem[]{}
   Shimura T., Takahara F. 1995, ApJ, 445, 780
 \bibitem[]{}
   Siemiginowska A., Czerny B.,  Kostyunin V., 1996, ApJ, 458, 491
 \bibitem[]{}
   Sobczak G. J., McClintock J. E.,  Remillard R. A., Cui W., Levine A. M.,
    Morgan E. H., Orosz J. A., Bailyn C. D., 2000, ApJ, 544, 993
 \bibitem[]{}
   Stella L., Rosner R., 1984, ApJ, 277, 312
 \bibitem[]{}
   Stern B.,  Poutanen J.,  Svensson R., Sikora M.,  Begelman M. C.,
       1995, ApJ, 449, L13
 \bibitem[]{}
   Titarchuk L. 1994, ApJ, 434, 570
 \bibitem[]{}
   Takizawa M. et al., 1997, ApJ, 489, 272
 \bibitem[]{}
   Tanaka Y., Lewin W. H. G.,  1995, in  Lewin W. H. G.,  van Paradijs J.,
     van den Heuvel E., eds, X--Ray Binaries. Cambridge Univ. Press, Cambridge,
     p. 126
 \bibitem[]{}
   Tanaka Y., Shibazaki N., 1996, ARA\&A, 34, 607
 \bibitem[]{}
   Turner M., J., L. et al.,   1989, PASJ, 41, 345
 \bibitem[]{}
   van der Klis M.,  1995, in  Lewin W. H. G.,  van Paradijs J.,
     van den Heuvel E., eds, X--Ray Binaries. Cambridge Univ. Press, Cambridge,
    p. 252
 \bibitem[]{}
   Watarai K., Fukue J., Takeuchi M.,  Mineshige S., 2000, PASP, 52, 133
 \bibitem[]{}
   Wilson C. D., Done C., 2001, MNRAS, in press, (astro-ph/0102167)
 \bibitem[]{}
   Witt H. J., Czerny B., \.{Z}ycki P. T., 1997, MNRAS, 286, 848
 \bibitem[]{}
   Zdziarski A. A., Johnson W. N., Magdziarz P., 1996, MNRAS, 283, 193
 \bibitem[]{}
   Zhang S. N., Cui W., Chen W., 1997a, ApJ, 482, L155
 \bibitem[]{}
   Zhang S. N., Cui W., Harmon B. A, Paciesas W. S., Remillard R. E.,
      van Paradijs J., 1997b, ApJ, 477, L95
 \bibitem[]{}
   Zhang S. N., Cui W., Chen  W., Yao Y., Zhang X., Sun X., Wu X.-B., Xu H.,
        2000, Science, 287, 1239
 \bibitem[]{}
   \.{Z}ycki P. T., Czerny B., 1994, MNRAS, 266, 653
 \bibitem[]{}
   \.{Z}ycki P. T., Done C.,  Smith D. A., 1997, ApJ, 488, L113 
 \bibitem[]{}
   \.{Z}ycki P. T., Done C.,  Smith D. A., 1998, ApJ, 496, L25 
 \bibitem[]{}
   \.{Z}ycki P. T., Done C.,  Smith D. A., 1999a, MNRAS, 305, 231
 \bibitem[]{}
   \.{Z}ycki P. T., Done C.,  Smith D. A., 1999b, MNRAS, 309, 561

\label{lastpage}

\end{thebibliography}
\end{document}